\documentclass[acmsmall,screen=true,natbib=true]{acmart_arxiv}

\usepackage{algorithm}
\usepackage{algorithmic}
\usepackage{eqparbox}

\usepackage{subfig}
\usepackage{xspace}
\usepackage{multirow}
\usepackage{setspace}
\usepackage{amsmath}
\usepackage{booktabs}
\DeclareMathOperator{\sign}{sgn}

\newcommand{\fnet}{F-FT-Net\xspace}
\newcommand{\fftnet}{F-FTT-Net\xspace}
\newcommand{\wnet}{W-FT-Net\xspace}
\newcommand{\wftnet}{W-FTT-Net\xspace}
\newcommand{\revise}[1]{\textcolor{black}{#1}}
\usepackage{wrapfig}

\AtBeginDocument{%
  \providecommand\BibTeX{{%
    \normalfont B\kern-0.5em{\scshape i\kern-0.25em b}\kern-0.8em\TeX}}}

\setcopyright{acmcopyright}

  \acmJournal{TODAES}
\acmYear{0000} \acmVolume{x} \acmNumber{x} \acmArticle{x} \acmMonth{1} \acmPrice{15.00}\acmDOI{xx.xxxx/xxxxxxx}

\begin{document}


\title{FTT-NAS: Discovering Fault-Tolerant Convolutional Neural Architecture}
\author{Xuefei~Ning}
\email{nxf16@mails.tsinghua.edu.cn}
\author{Guangjun~Ge}
\author{Wenshuo~Li}
\author{Zhenhua~Zhu}
\affiliation{\institution{Department of Electronic Engineering, Tsinghua University} \country{China}}
\author{Yin~Zheng}
\affiliation{\institution{Weixin Group, Tencent} \country{China}}
\author{Xiaoming~Chen}
\affiliation{\institution{State Key Laboratory of Computer Architecture, Institute of Computing Technology, Chinese Academy of Sciences} \country{China}}
\author{Zhen~Gao}
\affiliation{\institution{School of Electrical and Information Engineering, Tianjin University} \country{China}}
\author{Yu~Wang}
\email{yu-wang@tsinghua.edu.cn}
\author{Huazhong~Yang}
\email{yanghz@tsinghua.edu.cn}
\affiliation{\institution{Department of Electronic Engineering, Tsinghua University} \country{China}}

\renewcommand{\shortauthors}{X. Ning et al.}

\begin{abstract}
With the fast evolvement of embedded deep-learning computing systems, applications powered by deep learning are moving from the cloud to the edge. When deploying neural networks (NNs) onto the devices under complex environments, there are various types of possible faults: soft errors caused by cosmic radiation and radioactive impurities, voltage instability, aging, temperature variations, malicious attackers, etc. Thus the safety risk of deploying NNs is now drawing much attention. In this paper, after the analysis of the possible faults in various types of NN accelerators, we formalize and implement various fault models from the algorithmic perspective. We propose Fault-Tolerant Neural Architecture Search (FT-NAS) to automatically discover convolutional neural network (CNN) architectures that are reliable to various faults in nowadays devices. Then we incorporate fault-tolerant training (FTT) in the search process to achieve better results, which is referred to as FTT-NAS. Experiments on CIFAR-10 show that the discovered architectures outperform other manually designed baseline architectures significantly, with comparable or fewer floating-point operations (FLOPs) and parameters. Specifically, with the same fault settings, \fftnet discovered under the feature fault model achieves an accuracy of 86.2\% (VS. 68.1\% achieved by MobileNet-V2), and \wftnet discovered under the weight fault model achieves an accuracy of 69.6\% (VS. 60.8\% achieved by ResNet-18). By inspecting the discovered architectures, we find that the operation primitives, the weight quantization range, the capacity of the model, and the connection pattern have influences on the fault resilience capability of NN models.
\end{abstract}

\ccsdesc[500]{Hardware~Fault tolerance}
\ccsdesc[500]{Computing methodologies~Computer vision}

\keywords{neural architecture search, fault tolerance, neural networks}

\maketitle

\section{Introduction}
Convolutional Neural Networks (CNNs) have achieved breakthroughs in various tasks, including classification~\cite{resnet}, detection~\cite{ssd} and segmentation~\cite{long2015fully}, etc. Due to their promising performance, CNNs have been utilized in various safety-critic applications, such as autonomous driving, intelligent surveillance, and identification. Meanwhile, 
driven by the recent academic and industrial efforts, the neural network accelerators based on various hardware platforms (e.g., Application Specific Integrated Circuits (ASIC)~\cite{chen2014diannao}, Field Programmable Gate Array (FPGA)~\cite{qiu2016going}, Resistive Random-Access Memory (RRAM)~\cite{chi2016prime}) have been rapidly evolving.

The robustness and reliability related issues of deploying neural networks onto various embedded devices for safety-critical applications are attracting more and more attention. There is a large stream of algorithmic studies on various robustness-related characteristics of NNs, e.g., adversarial robustness~\cite{szegedy2013intriguing}, data poisoning~\cite{shafahi2018poison}, interpretability~\cite{zhang2018interpreting}, and so on. However, no hardware models are taken into consideration in these studies. Besides the issues from the purely algorithmic perspective, there exist hardware-related reliability issues when deploying NNs onto nowadays embedded devices. With the down-scaling of CMOS technology, circuits become more sensitive to cosmic radiation and radioactive impurities~\cite{henkel2013reliable}. Voltage instability, aging, and temperature variations are also common effects that could lead to errors. As for the emerging metal-oxide RRAM devices, due to the immature technology, they suffer from many types of device faults~\cite{chen2015rramdefect}, among which hard faults such as Stuck-at-Faults (SAFs) damage the computing accuracy severely and could not be easily mitigated~\cite{Xia2018StuckatFT}. Moreover, malicious attackers can attack the edge devices by embedding hardware Trojans, manipulating back-doors, and doing memory injection~\cite{zhao2019memory}.

Recently, some studies~\cite{liu2017rescuing,vialatte2017astudy,schorn2018accurate} analyzed the sensitivity of NN models. They proposed to predict whether a layer or a neuron is sensitive to faults and protect the sensitive ones. For fault tolerance, a straightforward way is to introduce redundancy in the hardware. Triple Modular Redundancy (TMR) is a commonly used but expensive method to tolerate a single fault~\cite{bolchini2007tmr,she2017reducing,zhao2019finegrained}. 
Studies~\cite{Xia2018StuckatFT,liu2017rescuing} proposed various redundancy schemes for Stuck-at-Faults tolerance in the RRAM-based Computing Systems. For increasing the algorithmic fault resilience capability, studies~\cite{he2019noise,hacene2019training} proposed to use fault-tolerant training (FTT), in which random faults are injected in the training process. 

Although redesigning the hardware for reliability is effective, it is not flexible and inevitably introduces a large overhead. It would be better if the issues could be mitigated as far as possible from the algorithmic perspective. Existing methods mainly concerned about designing training methods and analyzing the weight distribution~\cite{schorn2018accurate,he2019noise,hacene2019training}. Intuitively, the neural architecture might also be important for the fault tolerance characteristics~\cite{arechiga2018robustness,li2017understanding}, since it determines the ``path'' of fault propagation.
To verify these intuitions, the accuracies of baselines under a random bit-bias feature fault model\footnote{The random bit-bias feature fault model is formalized in Sec.~\ref{sec:feature-fault-model}.} are shown in Table~\ref{table:baseline-feature}, and the results under SAF weight fault model\footnote{The SAF weight fault model is formalized in Sec.~\ref{sec:weight-fault-model}.} are shown in Table~\ref{table:baseline-weights}. These preliminary experiments on the CIFAR-10 dataset show that the fault tolerance characteristics vary among neural architectures, which motivates the employment of neural architecture search (NAS) techniques in designing fault-tolerant neural architectures. We emphasize that our work is orthogonal to most of the previous methods based on hardware or mapping strategy design. To our best knowledge, our work is the first to increase the algorithmic fault resilience capability by optimizing the NN architecture.

\begin{table}[ht]
\centering
\caption{Performance of the baseline models with random bit-bias feature faults. $0/10^{-5}/10^{-4}$ denotes the per-MAC fault rate}
\label{table:baseline-feature}
\begin{tabular}{c|ccc}
\hline
Model     & Acc($0/10^{-5}/10^{-4}$) & \#Params & \#FLOPs \\ \hline
ResNet-18                      & 94.7/63.4/10.0 & 11.2M    & 1110M   \\
VGG-16$^\dagger$         & 93.1/21.4/10.0 & 14.7M    & 626M    \\
MobileNet-V2                   & 92.3/10.0/10.0 & 2.3M     & 182M    \\ \hline
\end{tabular}
\begin{minipage}{0.9\textwidth}
{\small
$\dagger$: For simplicity, we only keep one fully-connected layer of VGG-16.
}
\end{minipage}
\end{table}

\begin{table}[ht]
\centering
\caption{Performance of the baseline models with SAF weight faults. 0/4\%/8\% denotes the sum of the SAF1 and SAF0 rates}
\label{table:baseline-weights}
\begin{tabular}{c|ccc}
\hline
Model                & Acc(0/4\%/8\%) & \#Params & \#FLOPs \\ \hline
ResNet-18            & 94.7/64.8/17.8  & 11.2M    & 1110M   \\
VGG-16               & 93.1/45.7/14.3  & 14.7M    & 626M    \\
MobileNet-V2         & 92.3/26.2/11.7  & 2.3M     & 182M    \\ \hline
\end{tabular}
\end{table}

In this paper, we employ NAS to discover fault-tolerant neural network architectures against feature faults and weight faults, and demonstrate the effectiveness by experiments. The main contributions of this paper are as follows.

\begin{itemize}
    \item We analyze the possible faults in various types of NN accelerators (ASIC-based, FPGA-based, and RRAM-based), and formalize the statistical fault models from the algorithmic perspective. After the analysis, we adopt the Multiply-Accumulate (MAC)-i.i.d Bit-Bias (MiBB) model and the arbitrary-distributed Stuck-at-Fault (adSAF) model in the neural architecture search for tolerating feature faults and weight faults, respectively.
    \item We establish a multi-objective neural architecture search framework. On top of this framework, we propose two methods to discover neural architectures with better reliability: \textbf{FT-NAS} (NAS with a fault-tolerant multi-objective), and \textbf{FTT-NAS} (NAS with a fault-tolerant multi-objective and fault-tolerant training (FTT)).
    \item We employ FT-NAS and FTT-NAS to discover architectures for tolerating feature faults and weight faults. The discovered architectures, \fftnet and \wftnet have comparable or fewer floating-point operations (FLOPs) and parameters, and achieve better fault resilience capabilities than the baselines. 
    With the same fault settings, \fftnet discovered under the feature fault model achieves an accuracy of 86.2\% (VS. 68.1\% achieved by MobileNet-V2), and \wftnet discovered under the weight fault model achieves an accuracy of 69.6\% (VS. 60.8\% achieved by ResNet-18). 
    The ability of \wftnet to defend against several other types of weight faults is also illustrated by experiments. 
    \item We analyze the discovered architectures, and discuss how the weight quantization range, the capacity of the model, and the connection pattern influence the fault resilience capability of a neural network.
\end{itemize}

The rest of this paper is organized as follows. The related studies and the preliminaries are introduced in Section~\ref{sec:related-work}. In Section~\ref{sec:fault-model}, we conduct comprehensive analysis on the possible faults and formalize the fault models. In Section~\ref{sec:nas}, we elaborate on the design of the fault-tolerant NAS system. Then in Section~\ref{sec:exp}, the effectiveness of our method is illustrated by experiments, and the insights are also presented. Finally, we discuss and conclude our work in Section~\ref{sec:discussion} and Section~\ref{sec:conclusion}.

\section{Related work and preliminary}
\label{sec:related-work}
\subsection{Convolutional Neural Network}
Usually, a convolutional neural network is constructed by stacking multiple convolution layers and optional pooling layers, followed by fully-connected layers. 
Denoting the input feature map (IFM), before-activation output feature map, output feature map (OFM, i.e. activations), weights and bias of $i$-th convolution layer as \revise{$x^{(i)}$, $f^{(i)}$, $y^{(i)}$, $W^{(i)}$, $b^{(i)}$}, the computation can be written as:
\revise{\begin{equation}
    \begin{aligned}
    f^{(i)} &= W^{(i)} \circledast x^{(i)} + b^{(i)}\\
    y^{(i)} &= g(f^{(i)}),
    \end{aligned}
\end{equation}}
where $\circledast$ is the convolution operator, $g(\cdot)$ is the activation function, for which the ReLU function ($g(x) = \max(x, 0)$) is the commonest choice. From now on, we omit the \revise{$(i)$ superscript} for simplicity.

\subsection{NN Accelerators and Fixed-point Arithmetic}
With dedicated data flow design for efficient neural network processing, FPGA-based NN accelerators could achieve at least 10x better energy efficiency than GPUs~\cite{qiu2016going,guo2019survey}. And ASIC-based accelerators could achieve even higher efficiency~\cite{chen2014diannao}.
Besides, 
RRAM-based computing systems are promising solutions for energy-efficient brain-inspired computing~\cite{chi2016prime}, due to their capability of performing matrix-vector-multiplications (MVMs) in memory.
Existing studies have shown RRAM-based Processing-In-Memory (PIM) architectures can enhance the energy efficiency by over $100\times$ compared with both GPU and ASIC solutions, as they can eliminate the large data movements of bandwidth-bounded NN applications~\cite{chi2016prime}. For the detailed and formal hardware architecture descriptions, we refer the readers to the references listed above.

Currently, fixed-point arithmetic units are implemented by most of the NN accelerators, as 1) they consume much fewer resources and are much more efficient than the floating-point ones~\cite{guo2019survey}; 2) NN models are proven to be insensitive to quantization~\cite{qiu2016going,hubara2017quantized}. Consequently, quantization is usually applied before a neural network model is deployed onto the edge devices. 
To keep consistent with the actual deploying scenario, our simulation incorporates 8-bit dynamic fixed-point quantization for the weights and activations. More specifically, independent step sizes are used for the weights and activations of different layers. Denoting the fraction length and bit-width of a tensor as $l$ and $Q$, the step size (resolution) of the representation is $2^{-l}$. For common CMOS platforms, in which complement representation is used for numbers, the representation range of both weights and features is 
\begin{equation}
 [- 2^{Q-l}, 2^{-l} (2^{Q} - 1)].
\end{equation}

As for RRAM-based NN platforms, two separate crossbars are usually used for storing positive and negative weights~\cite{chi2016prime}. Thus the representation range of the weights (denoted by the $w$ superscript) is 
\begin{equation}
    [-R^w, R^w] = [- 2^{-l} (2^{Q+1} - 1), 2^{-l} (2^{Q+1} - 1)].
    \label{eq:rram_bound}
\end{equation}

For the feature representation in RRAM-based platforms, by assuming that the Analog to Digital Converters (ADCs) and Digital to Analog Converters (DACs) have enough precision, and the CMOS bit-width is $Q$-bit, the representation range of features (denoted by the $f$ superscript) in CMOS circuits is
\begin{equation}
    [-R^f, R^f] = [- 2^{Q-l}, 2^{-l} (2^{Q} - 1)].
    \label{eq:rram_bound_f}
\end{equation}

\subsection{Fault Resilience for CMOS-based Accelerators}

\citet{henkel2013reliable,borkar2005designing,slayman2011soft} revealed that advanced nanotechnology makes circuits more vulnerable to soft errors. Unlike hard errors, soft errors do not damage the underlying circuits, but instead trigger an upset of the logic state. The dominant cause of soft errors in CMOS circuits is the radioactive events, in which a single particle strikes an electronic device. \citet{arechiga2018robustness,libano2018selective} explored how the Single-Event Upset (SEU) faults impact the FPGA-based CNN computation system.

TMR is a commonly used approach to mitigate SEUs~\cite{bolchini2007tmr,she2017reducing,zhao2019finegrained}. 
Traditional TMR methods are agnostic of the NN applications and introduce large overhead. To exploit the NN applications' characteristics to reduce the overhead, one should understand the behavior of NN models with computational faults. \citet{vialatte2017astudy} analyzed the layer-wise sensitivity of NN models under two hypothetical feature fault models. \citet{libano2018selective} proposed to only triplicate the vulnerable layers after layer-wise sensitivity analysis and reduced the LUTs overhead for an NN model on Iris Flower from about 200\% (TMR) to 50\%. \citet{schorn2018accurate} conducted sensitivity analysis on the individual neuron level. \citet{li2017understanding} found that the impacts and propagation of computational faults in an NN computation system depend on the hardware data path, the model topology, and the type of layers. 
These methods analyzed the sensitivity of existing NN models at different granularities and exploited the resilience characteristics to reduce the hardware overhead for reliability. Our methods are complementary and discover NN architectures with better algorithmic resilience capability. 

To avoid the accumulation of the persistent soft errors in FPGA configuration registers, the scrubbing technique is applied by checking and partially reloading the configuration bits~\cite{bolchini2007tmr,xilinx2000partial}. 
From the algorithmic perspective, \citet{hacene2019training} demonstrated the effectiveness of fault-tolerant training in the presence of SRAM bit failures. 

\subsection{Fault Resilience for RRAM-based Accelerators}
RRAM devices suffer from lots of device faults~\cite{chen2015rramdefect}, among which the commonly occurring SAFs are shown to cause severe degradation in the performance of mapped neural networks~\cite{Xia2018StuckatFT}.
RRAM cells containing SAF faults get stuck at high-resistance state (SAF0) or low-resistance state (SAF1), thereby causing the weight to be stuck at the lowest or highest magnitudes of the representation range, respectively. Besides the hard errors, resistance programming variation~\cite{le2019resistive} is another source of faults for NN applications~\cite{liu2015vortex}.

For the detection of SAFs, \citet{Kannan2015Modeling,Kannan2013Sneak} proposed fault detection methods that can provide high fault coverage, \citet{xia2017fault} proposed on-line fault detection method that can periodically detect the current distribution of faults.

Most of the existing studies on improving the fault resilience ability of RRAM-based neural computation system focus on designing the mapping and retraining methods.  \citet{Xia2018StuckatFT,liu2017rescuing,xia2017fault,chen2017acceleratorfriendly} proposed different mapping strategies and the corresponding hardware redundancy design. After the distribution detection of the faults and variations, they proposed to retrain (i.e. finetune) the NN model for tolerating the detected faults, which is exploiting the intrinsic fault resilience capability of NN models. 
To overcome the programming variations, \citet{liu2015vortex} calculated the calibrated programming target weights with the log-normal resistance variation model, and proposed to map sensitive synapses onto cells with small variations. 
From the algorithmic perspective, \citet{liu2019afault} proposed to use error-correcting output codes (ECOC) to improve the NN's resilience capability for tolerating resistance variations and SAFs.

\subsection{Neural Architecture Search}
\label{sec:pre-nas}
Neural Architecture Search, as an automatic neural network architecture design method, has been recently applied to design model architectures for image classification and language models~\cite{nasnet,enas,DARTS}. The architectures discovered by NAS techniques have demonstrated surpassing performance than the manually designed ones. NASNet~\cite{nasnet} used a recurrent neural network (RNN) controller to sample architectures, trained them, and used the final validation accuracy to instruct the learning of the controller. Instead of using reinforcement learning (RL)-learned RNN as the controller, \citet{DARTS} used a relaxed differentiable formulation of the neural architecture search problem, and applied gradient-based optimizer for optimizing the architecture parameters; \citet{real2019aging} used evolutionary-based methods for sampling new architectures, by mutating the architectures in the population; Recent predictor-based search strategies~\cite{nao2018,ning2020generic} sample architectures with promising performance predictions, by gradient-based method~\cite{nao2018} or discrete inner search methods~\cite{ning2020generic}. Besides the improvements on the search strategies,
a lot of methods are proposed to speed up the performance evaluation in NAS. \citet{baker2017accelerating} incorporated learning curve extrapolation to predict the final performance after a few epochs of training; \citet{real2019aging} sampled architectures using mutation on existing models and initialized the weights of the sampled architectures by inheriting from the parent model; \citet{enas} shared the weights among different sampled architectures, and used the shared weights to evaluate each sampled architecture.

The goal of the NAS problem is to discover the architecture that maximizes some predefined objectives. The process of the original NAS algorithm goes as follows. At each iteration, $\alpha$ is sampled from the architecture search space $\mathcal{A}$. This architecture is then assembled as a candidate network $\mbox{Net}(\alpha,w)$, where $w$ is the weights to be trained. After training the weights $w$ on the training data split $D_t$, the evaluated reward of the candidate network on the validation data split $D_v$ will be used to instruct the sampling process. In its purest form, the NAS problem can be formalized as:
\begin{equation}\label{eq:nas_vanilla}
    \begin{aligned}
&\mbox{max}_{\alpha \in \mathcal{A}}\quad E_{x_v \sim D_v} [R(x_v, \mbox{Net}(\alpha,w^*(\alpha)))]\\
&\mbox{s.t. } w^*(\alpha) = \mbox{argmin}_{w} E_{x_t \sim D_t} [L(x_t, \mbox{Net}(\alpha,w))],
\end{aligned}
\end{equation}
where \revise{$\mathcal{A}$ is the architecture search space.} $\sim$ is the sampling operator, \revise{and $x_t, x_v$ denote the data sampled from the training and validation data splits $D_t, D_v$, respectively.} $E_{x \sim D}[\cdot]$ denotes the expectation with respect to the data distribution $D$, $R$ denotes the evaluated reward used to instruct the sampling process, and $L$ denotes the loss criterion for back propagation during the training of the weights $w$.

Originally, for the performance evaluation of each sampled architecture $\alpha$, one needs to find the corresponding $w^*(\alpha)$ by fully training the candidate network from scratch. This process is extremely slow, and shared weights evaluation is commonly used for accelerating the evaluation. In shared weights evaluation, each candidate architecture $\alpha$ is a subgraph of a super network and is evaluated using a subset of the super network weights. The shared weights of the super network are updated along the search process.

\section{Fault Models}
\label{sec:fault-model}

In Sec.~\ref{sec:fault-taxonomy}, we motivate and discuss the formalization of application-level statistical fault models.
Platform-specific analysis are conducted in Sec.~\ref{sec:fault-analysis-cmos} and Sec.~\ref{sec:fault-analysis-rram}. Finally, the MAC-i.i.d Bit-Bias (MiBB) feature fault model and the arbitrary-distributed Stuck-at-Fault model (adSAF) weight fault model are described in Sec.~\ref{sec:feature-fault-model} and Sec.~\ref{sec:weight-fault-model}, which would be used in the neural architecture search process. The analyses in this part are summarized in Fig.~\ref{fig:nas_framework} (a) and Table~\ref{table:fault-model-summary}.

\subsection{Application-Level Modeling of Computational Faults}
\label{sec:fault-taxonomy}
Computational faults do not necessarily result in functional errors~\cite{henkel2013reliable,li2017understanding}. For example, a neural network for classification tasks usually outputs a class probability vector, and our work only regards it as a functional error i.f.f the top-1 decision becomes different from the golden result. 
Due to the complexity of the NN computations and different functional error definition, it's very inefficient to incorporate gate-level fault injection or propagation analysis into the training or architecture search process. Therefore, to evaluate and further boost the algorithmic resilience of neural networks to computational faults, the application-level fault models should be formalized. 

From the algorithmic perspective, the faults fall into two categories: weight faults and feature faults. In this section, we analyze the possible faults in various types of NN accelerators, and formalize the statistical feature and weight fault models. A summary of these fault models is shown in Table~\ref{table:fault-model-summary}.

Note that we focus on the computational faults along the datapath inside the NN accelerator that could be modeled and mitigated from the algorithmic perspective. Faults in the control units and other chips in the system are not considered. See more discussion in the ``limitation of application-level fault models'' section in Sec.~\ref{sec:discuss_fault_model}.

\begin{table*}[tb]
\centering
\caption{Summary of the NN application-level statistical fault models, due to various types of errors on different platforms. \textbf{Headers}: H/S refers to Hard/Soft errors; P/T refers to Persistent/Transient influences; F/W refers to Feature/Weight faults}
\label{table:fault-model-summary}
 \resizebox{1.0\textwidth}{!}{
\begin{tabular}{c|c|c|c|c|c|c|c|c}
\toprule
\multirow{2}{*}{Platform} & \multirow{2}{*}{\begin{tabular}[c]{@{}c@{}}Error source\end{tabular}} & \multirow{2}{*}{\begin{tabular}[c]{@{}c@{}}Error\\position\end{tabular}}  & \multirow{2}{*}{\begin{tabular}[c]{@{}c@{}}Logic \\ component        \end{tabular}} & \multirow{2}{*}{\begin{tabular}[c]{@{}c@{}}H/S\end{tabular}}       & \multirow{2}{*}{\begin{tabular}[c]{@{}c@{}}P/T\end{tabular}} & \multirow{2}{*}{\begin{tabular}[c]{@{}c@{}}Common\\mitigation\end{tabular}} & \multicolumn{2}{c}{NN application level} \\ \cline{8-9} 
& & & & & & & \begin{tabular}[c]{@{}c@{}}F/W\end{tabular} &Simplified statistical model \\
\midrule
\hline

\multirow{2}{*}{RRAM}   & \multirow{2}{*}{SAF}        & SB-cell & \multirow{2}{*}{Crossbar} & \multirow{2}{*}{H} & \multirow{2}{*}{P}  & \multirow{2}{*}{\begin{tabular}[c]{@{}c@{}}detection+3R\\\cite{xia2017fault,liu2017rescuing,chen2017acceleratorfriendly} \end{tabular}} & \multirow{2}{*}{W}  & $w \sim \mbox{1bit-adSAF}(w_0; p_0, p_1)$ \\
\cline{3-3}\cline{9-9} & & MB-cell & & & & & & $w \sim Q\mbox{bit-adSAF}(w_0; p_0, p_1)$\\\hline

RRAM & \begin{tabular}[c]{@{}c@{}}variations\end{tabular} & MB-cell & Crossbar & S & P & \begin{tabular}[c]{@{}c@{}}PS loop\\\cite{liu2015vortex,hu2013bsb}\end{tabular} & W & \begin{tabular}[c]{@{}c@{}}$w \sim \mbox{LogNormal}(w_0; \sigma)$,\\ $w \sim \mbox{ReciprocalNormal}(w_0; \sigma)$\end{tabular} \\ \hline

\multirow{2}{*}{FPGA/ASIC} & SEE, overstress & \multirow{2}{*}{SRAM} & \multirow{2}{*}{Weight buffer}  & H & P & \multirow{2}{*}{ECC} & \multirow{2}{*}{W} & $w \sim \mbox{iBF}(w_0; r_h \times M_p(t))$\\
\cline{2-2}\cline{5-6}\cline{9-9} & \begin{tabular}[c]{@{}c@{}}SEE, VS \end{tabular} & & & S & T & & & $w \sim \mbox{iBF}(w_0; r_s)$\\ \hline \hline

\multirow{3}{*}{FPGA} & SEE, overstress & \multirow{3}{*}{LUTs} & \multirow{3}{*}{PE} & H & \multirow{3}{*}{P} & \begin{tabular}[c]{@{}c@{}}TMR\\\cite{bolchini2007tmr,she2017reducing,zhao2019finegrained}\end{tabular} & \multirow{3}{*}{F} & $f\sim \mbox{iBB}(f_0; r_h \times M_l \times M_p(t))$\\
\cline{2-2}\cline{5-5}\cline{7-7}\cline{9-9} & SEE, VS  & & & S & & \begin{tabular}[c]{@{}c@{}}TMR,\\Scrubbing~\cite{xilinx2000partial}\end{tabular} & &$f\sim \mbox{iBB}(f_0; r_s \times M_l \times M_p(t))$\\ \hline

\multirow{2}{*}{FPGA/ASIC/RRAM} & SEE, overstress & \multirow{2}{*}{SRAM} & \multirow{2}{*}{Feature buffer} & H & P & \multirow{2}{*}{ECC} & \multirow{2}{*}{F} & $y\sim \mbox{iBF}(y_0; r_h \times M_p(t))$\\
\cline{2-2}\cline{5-6}\cline{9-9}  & SEE, VS & & & S & T & & & $y\sim \mbox{iBF}(y_0; r_s)$\\ \hline

\multirow{2}{*}{ASIC} & SEE, overstress & \multirow{2}{*}{\begin{tabular}[c]{@{}c@{}}CL gates,\\ flip-flops\end{tabular}} & \multirow{2}{*}{PE} & H & P & \multirow{2}{*}{\begin{tabular}[c]{@{}c@{}}TMR,\\ DICE~\cite{dice}\end{tabular}} & \multirow{2}{*}{F} & $f\sim \mbox{iBB}(f_0; r^l_h \times M_l \times M_p(t))$\\
\cline{2-2}\cline{5-6}\cline{9-9} & SEE, VS &  & & S & T & & & $f\sim \mbox{iBB}(f_0; r^l_s \times M_l)$\\
\bottomrule
\end{tabular}
}
\begin{minipage}{1.0\textwidth}
{\small
\textbf{Notations}: $w, f, y$ refer to the weights, before-activation features, and after-activation features of a convolution; $p_0, p_1$ refer to the SAF0 and SAF1 rates of RRAM cells; $\sigma$ refers to the standard deviation of RRAM programming variations; $r_s, r_h$ refer to the soft and hard error rates of memory elements, respectively; $r_s^l, r_h^l$ refer to the soft and hard error rates of logic elements, respectively; $M_l$ is an amplifying coefficient for feature error rate due to multiple involved computational components; $M_p(t)>1$ is a coefficient that abstracts the error accumulation effects over time.

\textbf{Abbreviations}: SEE refers to Single-Event Errors, including Single-Event Burnout (SEB), Single-Event Upset (SEU), etc.; ``overstress'' includes conditions such as high temperature, voltage or physical stress; VS refers to voltage (down)scaling that is used for energy efficiency; SB-cell and MB-cell refer to single-bit and multi-bit memristor cells, respectively; CL gates refer to combinational logic gates; 3R refers to various Redundancy schemes and corresponding Remapping/Retraining techniques; PS loop refers to the programming-sensing loop during memristor programming; TMR refers to Triple Modular Redundancy; DICE refers to Dual Interlocked Cell.
}
\end{minipage}
\end{table*}

\subsection{Analysis of CMOS-based Platforms: ASIC and FPGA}
\label{sec:fault-analysis-cmos}

\begin{figure}[ht]
\begin{center}
\includegraphics[width=0.6\linewidth]{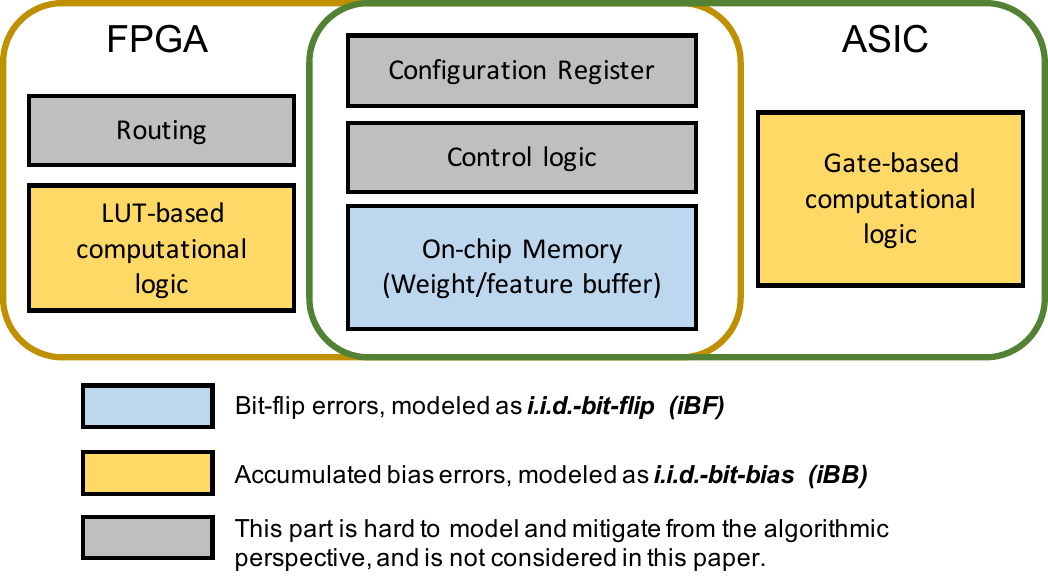}
\caption{The possible error positions in CMOS-based platforms.}
\label{fig:cmos_errors}
\end{center}
\end{figure}

The possible errors in CMOS-based platforms are illustrated in Fig.~\ref{fig:cmos_errors}. Soft errors that happen in the memory elements or the logic elements could lead to transient faulty outputs in ASICs. Compared with logic elements (e.g., combinational logic gates, flip-flops), memory elements are more susceptible to soft errors~\cite{slayman2011soft}. An unprotected SRAM cell usually has a larger bit soft error rate (SER) than flip-flops. 
Since the occurring probability of hard errors is much smaller than that of the soft errors, we focus on the analysis of soft errors, despite that hard errors lead to permanent failures.

The soft errors in the weight buffer could be modeled as i.i.d weight random bit-flips. Given the original value as \revise{$x_0$}, the distribution of a faulty value $x'$ under the random bit-flip (BF) model could be written as
\begin{equation}
\begin{aligned}
x' &\sim \mbox{BF}(x_0; p)\\
\mbox{indicates}\quad x' &= 2^{-l}(2^l x_0 \oplus e),\quad e = \sum_{q=1}^Q e_q 2^{q-1}\\
e_q &\sim \mbox{Bernoulli}(p),\quad q = 1, \cdots, Q,
\end{aligned}
\label{eq:bf_model}
\end{equation}
where $e_q$ denotes whether a bit-flip occurs at bit position $q$, $\oplus$ is the XOR operator.

By assuming that error occurs at each bit with an i.i.d bit SER of $r_s$, we know that each $Q$-bit weight has an i.i.d probability $p_w$ to encounter error, and $p_w = (1-(1-r_s)^Q) \approx r_s \times Q$, as $r_s \times Q \ll 1$. It is worth noting that throughout the analysis, we assume that the SERs of all components $\ll 1$, hence the error rate at each level is approximated as the sum of the error rates of the independent sub-components. As each weight encounters error independently, a weight tensor is distributed as i.i.d random bit-flip (iBF): $w \sim \mbox{iBF}(w_0; r_s)$, where $w_0$ is the golden weights. \citet{reagen2019ares} showed that the iBF model could capture the bit error behavior exhibited by real SRAM hardware. 

The soft errors in the feature buffer are modeled similarly as i.i.d random bit-flips, with a fault probability of approximately $r_s \times Q$ for $Q$-bit feature values. The distribution of the output feature map (OFM) values could be written as $y \sim \mbox{iBF}(y_0; r_s)$, where $y_0$ is the golden results.

Actually, FPGA-based implementations are usually more vulnerable to soft errors than their ASIC counterparts~\cite{asadi2007analytical}. Since the majority space of an FPGA chip is filled with memory cells, the overall SER rate is much higher. Moreover, the soft errors occurring in logic configuration bits would lead to persistent faulty computation, rather than transient faults as in ASIC logic. Persistent errors can not be mitigated by simple retry methods and would lead to statistically significant performance degradation. Moreover, since the persistent errors would be accumulated if no correction is made, the equivalent error rate would keep increasing as time goes on. We abstract this effect with a monotonic increasing function $M_p(t) \geq 1$, where the subscript $p$ denotes ``persistent'', and $t$ denotes the time. \revise{For example, if the FPGA weight buffer or LUTs are reloaded for every $T$ period in the radioactive environment~\cite{bolchini2007tmr,xilinx2000partial}, a multiplier of $M_p(T)$ would be the worst bounding case. Note that the exact choice of $t$ is not important in our experiments, since our work mainly aims at comparing different neural architectures using certain fault insertion pattern and ratio, and the temporal effect modeled by $M_p(t)$ does not influence the architectural preference.}

Let us recap how one convolution is mapped onto the FPGA-based accelerator, to see what the configuration bit errors could cause on the OFM values. If the dimension of the convolution kernel is $(c, k, k)$ (channel, kernel height, kernel width, respectively), there are $c k^2-1 \approx c k^2$ additions needed for the computation of a feature value. We assume that the add operations are spatially expanded onto adder trees constructed by LUTs, i.e., no temporal reusing of adders is used for computing one feature value. That is to say, the add operations are mapped onto different hardware adders\footnote{See more discussion in Sec.~\ref{sec:discuss_fault_model}.}, and encounter errors independently. 
The per-feature error rate could be approximated by the adder-wise SER times $M_l$, where $M_l \approx c k^2$. 
Now, let's dive into the adder-level computation, in a 1-bit adder with scale $s$, the bit-flip in one LUTs bit would add a bias $\pm 2^s$ to the output value, if the input bit signals match the address of this LUTs bit.
If each LUT cell has an i.i.d SER of $r_s$, in a $Q'$-bit adder, denoting the fraction length of the operands and result as $l'$, the distribution of the faulty output $x'$ with the random bit-bias (BB) faults could be written as
\begin{equation}
\begin{aligned}
x' &\sim \mbox{BB}(x_0; p, Q', l')\\
\mbox{indicates}\quad x' &= x_0 + e,\quad e = 2^{-l'} \sum_{q=1}^{Q'} (-1)^\beta 2^{q-1} e_q \\
e_q &\sim \mbox{Bernoulli}(p)\\
\beta_q &\sim \mbox{Bernoulli}(0.5),\quad q = 1, \cdots, Q'.
\end{aligned}
\label{eq:bb_model}
\end{equation}

As for the result of the adder tree constructed by multiple LUT-based adders, since the probability that multiple bit-bias errors co-occur is orders of magnitude smaller, we ignore the accumulation of the biases that are smaller than the OFM quantization resolution $2^{-l}$. Consequently, the OFM feature values before the activation function follow the i.i.d Random Bit-Bias distribution $f \sim \mbox{iBB}(f_0; r_s \times M_l \times M_p(t), Q, l)$, where $Q$ and $l$ are the bit-width and fraction length of the OFM values, respectively.


We can make an intuitive comparison between the equivalent feature error rates induced by LUTs soft errors and feature buffer soft errors. As the majority of FPGAs is SRAM-based, considering the bit SER $r_s$ of LUTs cell and BRAM cell to be close, we can see that the feature error rate induced by LUTs errors is amplified by $M_l \times M_p(t)$. As we have discussed, $M_p(t) \geq 1, M_l = c k^2 > 1$, 
the performance degradation induced by LUTs errors could be significantly larger than that induced by feature buffer errors.

\subsection{Analysis of PIM-based Platforms: RRAM as an example}
\label{sec:fault-analysis-rram}

In an RRAM-based computing system, compared with the accompanying CMOS circuits, the RRAM crossbar is much more vulnerable to various non-ideal factors. In multi-bit RRAM cells, studies have showed that the distribution of the resistance due to programming variance is either Gaussian or Log-Normal~\cite{le2019resistive}. As each weight is programmed as the conductance of the memristor cell, the weight could be seen as being distributed as Reciprocal-Normal or Log-Normal. Besides the soft errors, common hard errors such as SAFs, caused by fabrication defects or limited endurance, could result in severe performance degradation~\cite{Xia2018StuckatFT}. SAFs occur frequently in nowadays RRAM crossbar: As reported by \cite{chen2015rramdefect}, the overall SAF ratio could be larger than 10\% ($p_1=9.04\%$ for SAF1 and $p_0=1.75\%$ for SAF0) in a fabricated RRAM device. The statistical model of SAFs in single-bit and multi-bit RRAM devices would be formalized in Sec.~\ref{sec:weight-fault-model}.

As the RRAM crossbars also serve as the computation units, some non-ideal factors (e.g., IR-drop, wire resistance) could be abstracted as feature faults. They are not considered in this work since the modeling of these effects highly depends on the implementation (e.g., crossbar dimension, mapping strategy) and hardware-in-the-loop testing~\cite{he2019noise}.

\subsection{Feature Fault Model}
\label{sec:feature-fault-model}
As analyzed in Sec.~\ref{sec:fault-analysis-cmos}, the soft errors in LUTs are relatively the more pernicious source of feature faults, as 1) SER is usually much higher than hard error rate: $r_s \gg r_h$, 2) these errors are persistent if no correction is made, 3) the per-feature equivalent error rate is amplified as multiple adders are involved. 
Therefore, we use the iBB fault model in our exploration of mitigating feature faults.

\begin{figure}[htb]
\begin{center}
\includegraphics[width=0.5\linewidth]{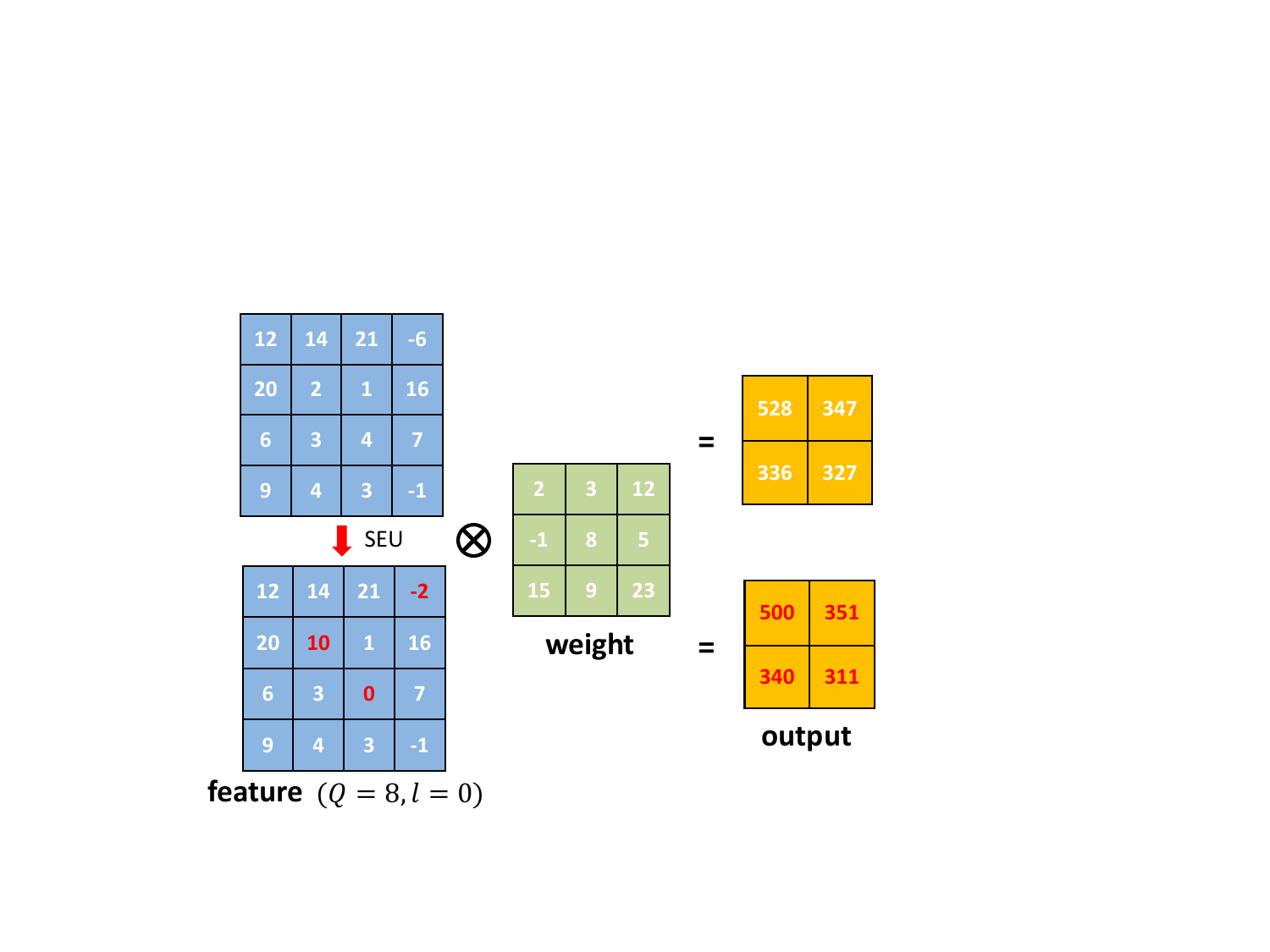}
\caption{An example of injecting feature faults \revise{under the iBB fault model (soft errors in FPGA LUTs).}}
\label{fig:feature_fault_inject}
\end{center}
\end{figure}

We have $f \sim \mbox{iBB}(f_0; r_s M_l M_p(t))$, where $M_l = ck^2$, and the probability of error occurring at every position in the OFM is $p=r_s M_l M_p(t) Q = p_m M_l$, where $p_m=r_s Q M_p(t)$ is defined as the per-MAC error rate. Denoting the dimension of the OFM as $(C_o, H, W)$ (channel, height, and width, respectively) and the dimension of each convolution kernel as $(c, k, k)$, the computation of a convolution layer under this fault model could be written as
\begin{equation}
\begin{aligned}
y &= g(W \circledast x + b + \theta \cdot 2^{\alpha-l} \cdot (-1)^{\beta})\\
\mbox{s.t.}\quad \theta &\sim \mbox{Bernoulli}(p)^{C_o\times H \times W}\\
\alpha &\sim U\{0, \cdots, Q-1\}^{C_o\times H \times W}\\
\beta &\sim U\{0,1\}^{C_o\times H \times W},
\end{aligned}
\end{equation}
where $\theta$ is the mask indicating whether an error occurs at each feature map position, $\alpha$ represents the bit position of the bias, $\beta$ represents the bias sign. Note that this formulation is not equivalent to the random bit-bias formalization in Eq.~\ref{eq:bb_model}, and is adopted for efficient simulation. These two formulations are close when the odds that two errors take effect simultaneously is small ($p_m / Q \ll 1$). 
This fault model is referred to as the MAC-i.i.d Bit-Bias model (abbreviated as MiBB). An example of injecting this type of feature faults is illustrated in Fig.~\ref{fig:feature_fault_inject}. 

Intuitively, convolution computation that needs fewer MACs might be more immune to the faults, as the equivalent error rate at each OFM location is lower.

\subsection{Weight Fault Model}
\label{sec:weight-fault-model}

As RRAM-based accelerators suffer from a much higher weight error rate than the CMOS-based ones. The Stuck-at-Faults in RRAM crossbars are mainly considered for the setup of the weight fault model. 
We assume the underlying platform is RRAM with multi-bit cells, and adopt the commonly-used mapping scheme, in which separate crossbars are used for storing positive and negative weights~\cite{chi2016prime}. That is to say, when an SAF0 fault causes a cell to be stuck at HRS, the corresponding logical weight would be stuck at 0. When an SAF1 fault causes a cell to be stuck at LRS, the weight would be stuck at $-R^w$ if it's negative, or $R^w$ otherwise.

\begin{figure}[ht]
\begin{center}
\includegraphics[width=0.65\linewidth]{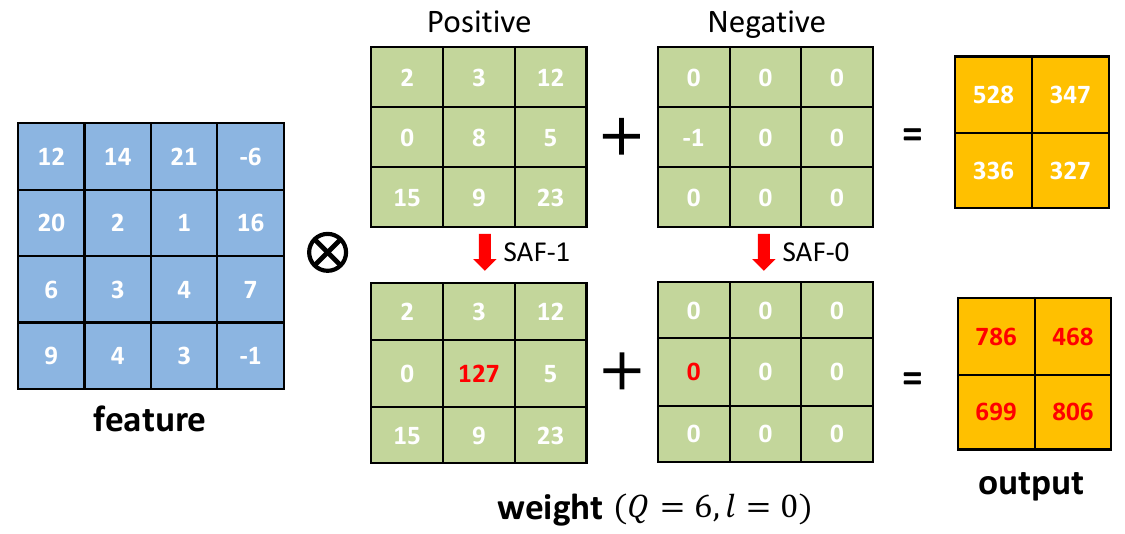}
\caption{An example of injecting weight faults \revise{under the adSAF fault model (SAF errors in RRAM cells).}}
\label{fig:weight_fault_inject}
\end{center}
\end{figure}

The computation of a convolution layer under the SAF weight fault model could be written as
\begin{equation}
    \begin{aligned}
    y &= g(W' \circledast x + b)\\
    \mbox{s.t.}\quad W' &= (1-\theta) \cdot W +  \theta \cdot e\\
    \theta &\sim \mbox{Bernoulli}(p_0 + p_1)^{C_o \times c \times k\times k }\\
    m &\sim \mbox{Bernoulli}(\frac{p_1}{p_0 + p_1})^{C_o\times c \times k\times k }\\
    e &= R^w \sign(W) \cdot m,
    \end{aligned}
    \label{eq:weight_fault_model}
\end{equation}
where $R^w$ refers to the representation bound in Eq.~\ref{eq:rram_bound}, $\theta$ is the mask indicating whether fault occurs at each weight position, $m$ is the mask representing the SAF types (SAF0 or SAF1) at faulty weight positions, $e$ is the mask representing the faulty target values ($0$ or $\pm R^w$). Every single weight has an i.i.d probability of $p_0$ to be stuck at $0$, and $p_1$ to be stuck at the positive or negative bounds of the representation range, for positive and negative weights, respectively. An example of injecting this type of weight faults is illustrated in Fig.~\ref{fig:weight_fault_inject}.

\begin{figure*}[ht]
\begin{center}
\includegraphics[width=0.98\linewidth]{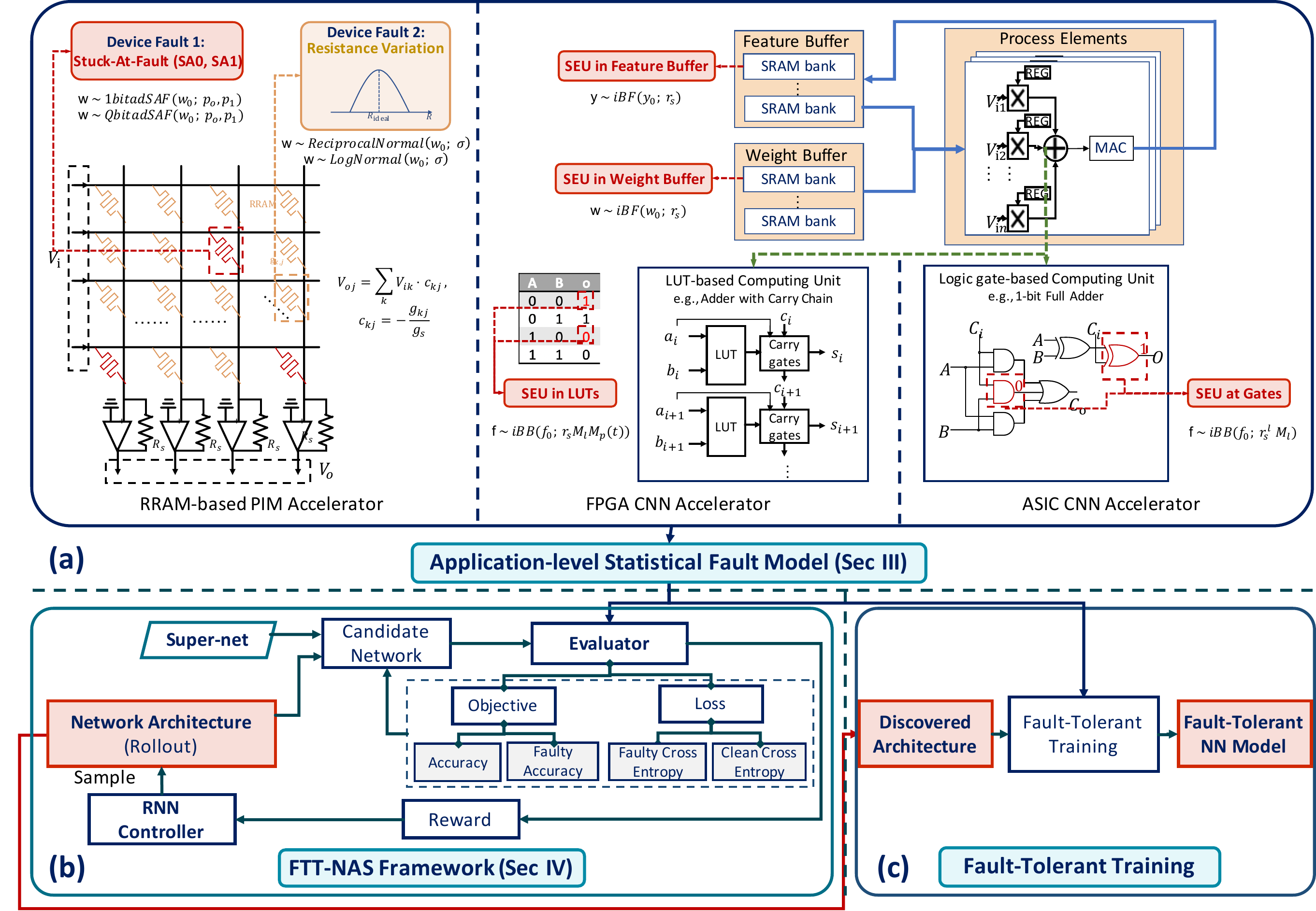}
\caption{Illustration of the overall workflow. (a) The setup of the application-level statistical fault models. (b) The FTT-NAS framework. (c) The final fault-tolerant training stage.}
\label{fig:nas_framework}
\end{center}
\end{figure*}

Note that the weight fault model, referred to as arbitrary-distributed Stuck-at-Fault model (adSAF), is much harder to defend against than SAF faults with a specific known defect map. A neural network model that behaves well
under the adSAF model is expected to achieve high reliability across different specific SAF defect maps.

The above adSAF fault model assumes the underlying hardware is multi-bit RRAM devices, adSAFs in single-bit RRAM devices are also of interest. In single-bit RRAM devices, multiple bits of one weight value are mapped onto different crossbars, of which the results would be shifted and added together~\cite{zhu2019aconfigurable}. In this case, an SAF fault that occurs in a cell would cause the corresponding bit of the corresponding weight to be stuck at $0$ or $1$. The effects of adSAF faults on a weight value in single-bit RRAM devices can be formulated as
\begin{equation}
\begin{aligned}
    w' &= \sign(w) 2^{-l}(((\neg \theta) \land 2^l |w|) \lor (\theta \land e))\\
    \theta &= \sum_{q=1}^Q \theta_q 2^{q-1},\quad e = \sum_{q=1}^Q m_q 2^{q-1}\\
    \theta_q &\stackrel{iid}{\sim} \mbox{Bernoulli}(p_0 + p_1),\quad q = 1, \cdots, Q\\
    m_q &\stackrel{iid}{\sim} \mbox{Bernoulli}(\frac{p_1}{p_0 + p_1}),\quad q = 1, \cdots, Q,
\end{aligned}
\end{equation}
where the binary representation of $\theta$ indicates whether fault occurs at each bit position, the binary representation of $e$ represents the target faulty values ($0$ or $1$) at each bit position if fault occurs. We will demonstrate that the architecture discovered under the multi-bits adSAF fault model can also defend against single-bit adSAF faults and iBF weight faults caused by errors in the weight buffers of CMOS-based accelerators.

\section{Fault-Tolerant NAS}
\label{sec:nas}
In this section, we present the FTT-NAS framework. We first give out the problem formalization and framework overview in Sec.~\ref{sec:nas-overview}. Then, the search space, sampling and assembling process are described in Sec.~\ref{sec:nas-ss} and Sec.~\ref{sec:nas-sanda}, respectively. Finally, the search process is elaborated in Sec.~\ref{sec:nas-search}.

\subsection{Framework Overview}
\label{sec:nas-overview}
Denoting the fault distribution characterized by the fault models as $F$, the neural network search for fault tolerance can be formalized as
\begin{equation}
\label{eq:nas}
    \begin{aligned}
&\mbox{max}_{\alpha \in \mathcal{A}}\quad E_{x_v \sim D_v} [E_{f\sim F}[R(x_v, \mbox{Net}(\alpha, w^*(\alpha)), f)]]\\
&\mbox{s.t. } w^*(\alpha) = \mbox{argmin}_{w} \: E_{x_t \sim D_t} [E_{f \sim F}[L(x_t, \mbox{Net}(\alpha, w), f)]].
\end{aligned}
\end{equation}
\revise{where $\mathcal{A}$ is the architecture search space, and $D_t, D_v$ denote the training and validation data split, respectively. $R$ and $L$ denote the reward and loss criterion, respectively. The major difference of Eq.~\ref{eq:nas} to the vanilla NAS problem Eq.~\ref{eq:nas_vanilla} lies in the introduction of the fault model $F$.}

As the cost of finding the best weights $w^*$ for each architecture $\alpha$ is almost unbearable, we use the shared-weights based evaluator, in which shared weights are directly used to evaluate sampled architectures. The resulting method, FTT-NAS, is the method to solve this NAS problem approximately. And FT-NAS can be viewed as a degraded special case for FTT-NAS, in which no fault is injected in the inner optimization of finding $w^*(\alpha)$.

The overall neural architecture search framework is illustrated in Fig.~\ref{fig:nas_framework} (b). There are multiple components in the framework: A \textbf{controller} that samples different architecture rollouts from the \textbf{search space}; A \textbf{candidate network} is assembled by taking out the corresponding subset of weights from the \textbf{super-net}. A shared weights based \textbf{evaluator} evaluates the performance of different rollouts on the CIFAR10 \textbf{dataset}, using fault-tolerant \textbf{objectives}.

\subsection{Search Space}
\label{sec:nas-ss}
\begin{figure}[ht]
\begin{center}
\includegraphics[width=0.45\linewidth]{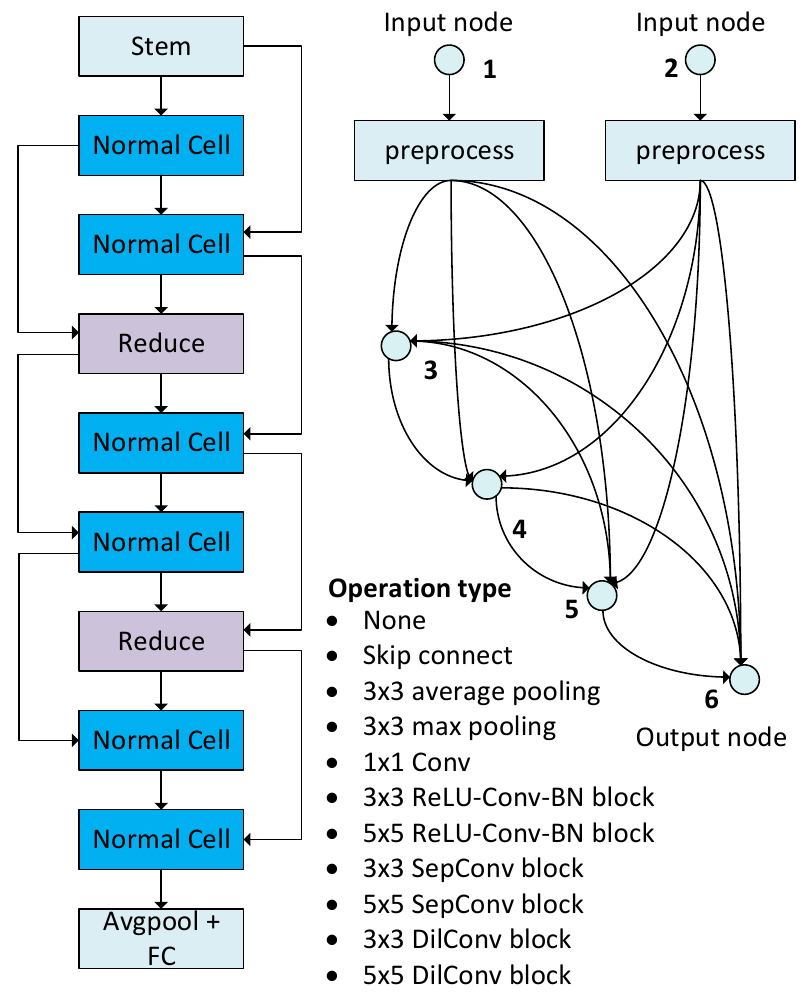}
\caption{Illustration of the search space design. Left: The layout and connections between cells. Right: The possible connections in each cell, and the possible operation types on every connection.}
\label{fig:search_space}
\end{center}
\end{figure}

The design of the search space is as follows: We use a cell-based macro architecture, similar to the one used in \cite{enas,DARTS}. There are two types of cells: normal cell, and reduction cell with stride 2. All normal cells share the same connection topology, while all reduction cells share another connection topology. The layout and connections between cells are illustrated in Fig.~\ref{fig:search_space}.

In every cell, there are $B$ nodes, node 1 and node 2 are treated as the cell's inputs, which are the outputs of the two previous cells.
For each of the other $B-2$ nodes, two in-coming connections will be selected and element-wise added. For each connection, the 11 possible operations are: none; skip connect; 3x3 average (avg.) pool; 3x3 max pool; 1x1 Conv; 3x3 ReLU-Conv-BN block; 5x5 ReLU-Conv-BN block; 3x3 SepConv block; 5x5 SepConv block; 3x3 DilConv block; 5x5 DilConv block.

The complexity of the search space can be estimated. For each cell type, there are $(11^{(B-2)}\times (B-1)!)^2$ possible choices. As there are two independent cell types, there are $(11^{(B-2)} \times (B-1)!)^4$ possible architecture in the search space, which is roughly $9.5 \times 10^{24}$ with $B=6$ in our experiments.

\subsection{Sampling and Assembling Architectures}
\label{sec:nas-sanda}
In our experiments, the controller is an RNN, and the performance evaluation is based on a super network with shared weights, as used by \cite{enas}.

\begin{figure}[tb]
\begin{center}
\begin{minipage}{6cm}
\includegraphics[width=0.8\linewidth]{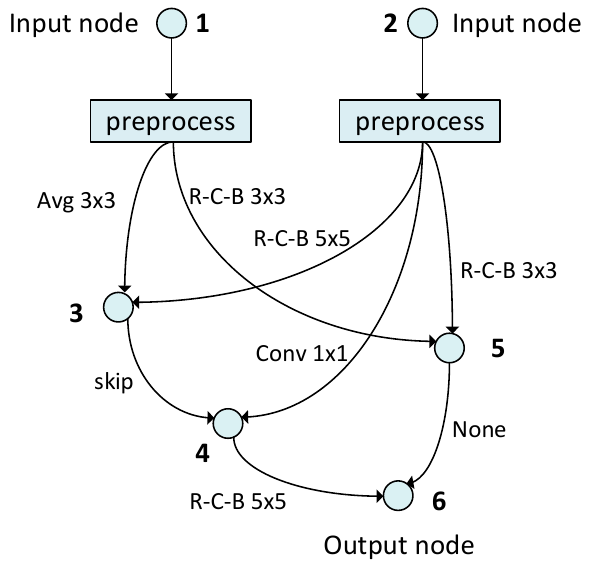}
\end{minipage}
\caption{An example of the sampled cell architecture.}
\label{fig:cell_example}
\end{center}
\end{figure}

An example of the sampled cell architecture is illustrated in Fig.~\ref{fig:cell_example}. Specifically, to sample a cell architecture, the controller RNN samples $B-2$ blocks of decisions, one for each node $3, \cdots, B$. In the decision block for node $i$, $M=2$ input nodes are sampled from $1, \cdots, i-1$, to be connected with node $i$. Then $M$ operations are sampled from the $11$ basic operation primitives, one for each of the $M$ connections. Note that the two sampled input nodes can be the same node $j$, which will result in two independent connections from node $j$ to node $i$. 

During the search process, the architecture assembling process using the shared-weights super network is straightforward~\cite{enas}: Just take out the weights from the super network corresponding to the connections and operation types of the sampled architecture.

\begin{algorithm}[ht]
\begin{algorithmic}[1]
\STATE EPOCH: the total search epochs
\STATE ${\bf w}$: shared weights in the super network
\STATE ${\bf \theta}$: the parameters of the controller \revise{$\pi$}

\STATE epoch = 0
\WHILE{epoch $<$ EPOCH}
\FORALL{$x_t, y_t \sim D_t$}
\STATE $a \sim \pi(a;\theta)$ \COMMENT{\revise{sample an architecture from the controller}}
\STATE $f \sim F(f)$ \COMMENT{\revise{sample faults from the fault model}}
\STATE $L_c = \mbox{CE}(\mbox{Net}(a; w)(x_t), y_t)$ \COMMENT{clean cross entropy}
\STATE $L_f = \mbox{CE}(\mbox{Net}(a; w)(x_t, f), y_t)$ \COMMENT{faulty cross entropy}
\STATE $L(x_t, y_t, \mbox{Net}(a; w), f) = (1-\alpha_l) L_c + \alpha_l L_f$
\STATE $w = w - \eta_w \nabla_w L$  \COMMENT{for clarity, we omit momentum calculation here}
\ENDFOR
\FORALL{$x_v, y_v \sim D_v$}
\STATE $a \sim \pi(a;\theta)$ \COMMENT{\revise{sample an architecture from the controller}}
\STATE $f \sim F(f)$ \COMMENT{\revise{sample faults from the fault model}}
\STATE $R_c = \mbox{Acc}(\mbox{Net}(a; w)(x_v), y_v)$ \COMMENT{clean accuracy}
\STATE $R_f = \mbox{Acc}(\mbox{Net}(a; w)(x_v, f), y_v)$\COMMENT{faulty accuracy}
\STATE $R(x_v, y_v, \mbox{Net}(a; w), f) = (1-\alpha_r) * R_c + \alpha_r * R_f$
\STATE $\theta = \theta + \eta_\theta (R - b) \nabla_\theta \log \pi(a;\theta)$
\ENDFOR
\STATE epoch = epoch + 1
\STATE schedule $\eta_w, \eta_\theta$
\ENDWHILE
\RETURN $a \sim \pi(a; \theta)$
\end{algorithmic}
\caption{FTT-NAS}
\label{alg:nas}
\end{algorithm}

\subsection{Searching for Fault-Tolerant Architecture}
\label{sec:nas-search}
The FTT-NAS algorithm is illustrated in Alg.~\ref{alg:nas}. To search for a fault-tolerant architecture, we use a weighted sum of the clean accuracy $\mbox{acc}_c$ and the accuracy with fault injection $\mbox{acc}_f$ as the reward to instruct the training of the controller:
\begin{equation}
\label{eq:reward}
    R = (1-\alpha_r) * \mbox{acc}_c + \alpha_r * \mbox{acc}_f,
\end{equation}
where $\mbox{acc}_f$ is calculated by injecting faults following the fault distribution described in Sec.~\ref{sec:fault-model}. For the optimization of the controller, we employ the Adam optimizer~\cite{kingma2015adam} to optimize the REINFORCE~\cite{williams1992simple} objective, together with an entropy encouraging regularization.

In every epoch of the search process, we alternatively train the shared weights and the controller on separate data splits $D_t$ and $D_v$, respectively. For the training of the shared weights, we carried out experiments under two different settings: without/with FTT. When training with FTT, a weighted sum of the clean cross entropy loss $\mbox{CE}_c$ and the cross entropy loss with fault injection $\mbox{CE}_f$ is used to train the shared weights. The FTT loss can be written as
\begin{equation}
\label{eq:loss}
    \emph{L} = (1-\alpha_l) * \mbox{CE}_c + \alpha_l * \mbox{CE}_f.
\end{equation}

As shown in line 7-12 in Alg.~\ref{alg:nas}, in each step of training the shared weights, we sample architecture $\alpha$ using the current controller, then backpropagate using the FTT loss to update the parameters of the candidate network. Training without FTT (in FT-NAS) is a special case with $\alpha_l = 0$.

As shown in line 15-20 in Alg.~\ref{alg:nas}, in each step of training the controller, we sample architecture from the controller, assemble this architecture using the shared weights, and then get the reward $R$ on one data batch in $D_v$. Finally, the reward is used to update the controller by applying the REINFORCE technique~\cite{williams1992simple}, with the reward baseline denoted as $b$.

\section{Experiments}
\label{sec:exp}

In this section, we demonstrate the effectiveness of the FTT-NAS framework and analyze the discovered architectures under different fault models. First, we introduce the experiment setup in Sec.~\ref{sec:exp-setup}. Then, the effectiveness under the feature and weight fault models are shown in Sec.~\ref{sec:exp-feature} and Sec.~\ref{sec:exp-weight}, respectively. The effectiveness of the learned controller is illustrated in Sec.~\ref{sec:exp-controller}. Finally, the analyses and illustrative experiments are presented in Sec.~\ref{sec:inspection}.

\subsection{Setup}
\label{sec:exp-setup}
Our experiments are carried out on the CIFAR-10~\cite{cifar10} dataset. CIFAR-10 is one of the most commonly used computer vision datasets and contains 60000 $32\times 32$ RGB images. Three manually designed architectures VGG-16, ResNet-18, and MobileNet-V2 are chosen as the baselines. 8-bit dynamic fixed-point quantization is used throughout the search and training process, and the fraction length is found following the minimal-overflow principle.

In the neural architecture search process, we split the training dataset into two subsets. 80\% of the training data is used to train the shared weights, and the remaining 20\% is used to train the controller. The super network is an 8-cell network, with all the possible connections and operations. The channel number of the first cell is set to 20 during the search process, and the channel number increases by 2 upon every reduction cell. 
The controller network is an RNN with one hidden layer of size 100. The learning rate for training the controller is 1e-3. The reward baseline $b$ is updated using a moving average with momentum 0.99.  To encourage exploration, we add an entropy encouraging regularization to the controller's REINFORCE objective, with a coefficient of $0.01$. 
For training the shared weights, we use an SGD optimizer with momentum 0.9 and weight decay 1e-4, the learning rate is scheduled by a cosine annealing scheduler~\cite{loshchilov2016sgdr}, started from $0.05$. Each architecture search process is run for 100 epochs. Note that all these are typical settings that are similar to \cite{enas}.

\revise{To conduct the final training of the architectures (Fig.\ref{fig:nas_framework} (c)), we run fault-tolerant training for 100 epochs. The learning rate is set to $0.1$ initially and decayed by $10$ at epoch 40 and 80. We have experimented with a fault-tolerant training choice: whether to mask out the error positions in feature/weights during the backpropagation process. If the error positions are masked out, then no gradient would be backpropagated through the erroneous feature positions, and no gradient would be calculated w.r.t. the erroneous weight positions. We find that this choice does not affect the fault-tolerant training result, thus we do not use the masking operation in our final experiments.}

We build the neural architecture search framework and fault injection framework upon the PyTorch framework, and all the codes are available at \url{https://github.com/walkerning/aw_nas}. 
 
\subsection{Defend Against MiBB Feature Faults}
\label{sec:exp-feature}

\begin{table*}[ht]
\centering
\caption{Comparison of different architectures under the MiBB feature fault model}
\label{table:NAS-f-results}
\begin{tabular}{cc||ccccccccc}
\hline
\multirow{2}{*}{Arch}  & \multirow{2}{*}{Training$^{\dagger}$} & \multirow{2}{*}{\begin{tabular}[c]{@{}c@{}}Clean\\ accuracy\end{tabular}} & \multicolumn{5}{c}{Accuracy (\%) with feature faults}    & \multirow{2}{*}{\#FLOPs} & \multirow{2}{*}{\#Params} \\ \cline{4-8}
                        & & & 3e-6  & 1e-5    & 3e-5    & 1e-4 & 3e-4&                          &                           \\ \hline
ResNet-18           & clean     & 94.7 & 89.1 & 63.4 & 11.5 & 10.0 & 10.0 & 1110M & 11.16M   \\
VGG-16              & clean     & 93.1 & 78.2 & 21.4 & 10.0 & 10.0 & 10.0 & 626M  & 14.65M   \\
MobileNet-V2        & clean     & 92.3 & 10.0  & 10.0 & 10.0 & 10.0 & 10.0 & 182M  & 2.30M    \\\hline
\textbf{\fnet}      & clean     & 91.0   & 71.3  & 22.8 & 10.0 & 10.0 & 10.0 & 234M &     0.61M \\
\hline \hline
ResNet-18           & $p_m$=1e-4  & 79.2 & 79.1 & 79.6 & 78.9 & 60.6 & 11.3 & 1110M & 11.16M    \\
VGG-16              & $p_m$=3e-5  & 83.5 & 82.4 & 77.9 & 50.7 & 11.1 & 10.0 & 626M  & 14.65M    \\                        
MobileNet-V2        & $p_m$=3e-4  & 71.2 & 70.3 & 69.0 & 68.7 & 68.1 & 47.8 & 182M  & 2.30M    \\\hline
\textbf{\fftnet}    & $p_m$=3e-4  & {\bf 88.6} & {\bf 88.7} & {\bf 88.5} & {\bf 88.0} & {\bf 86.2} & {\bf 51.0} & 245M  & 0.65M \\\hline
\end{tabular}
\begin{minipage}{0.96\textwidth}
{\small
$\dagger$: As also noted in the main text, for all the FTT trained models, we successively try per-MAC fault injection probability $p_m$ in \{3e-4, 1e-4, 3e-5\}, and use the largest injection probability with which the model could achieve a clean accuracy above 50\%.
}
\end{minipage}
\end{table*}

\begin{figure*}[ht]
\centering
\subfloat[]{\includegraphics[width=0.47\linewidth]{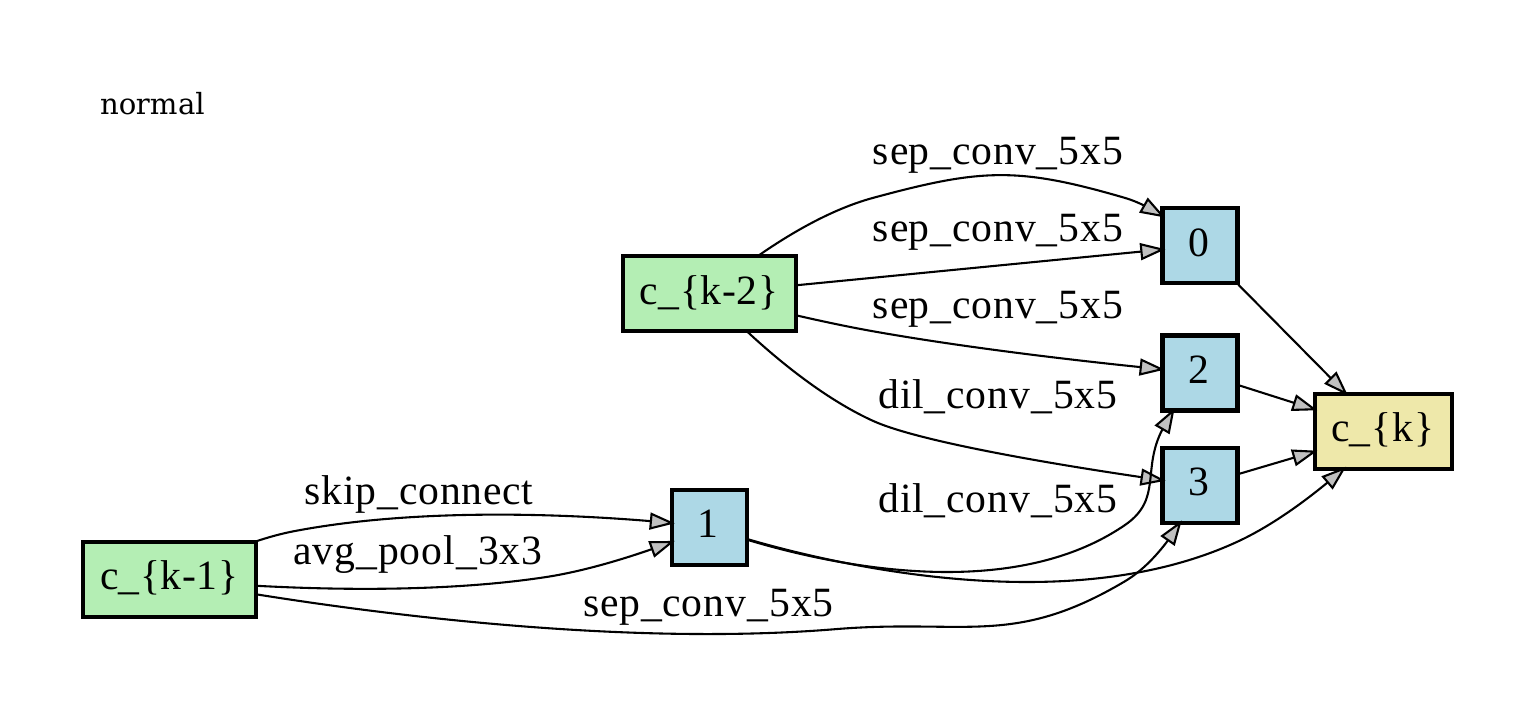}
}
\hfil
\subfloat[]{\includegraphics[width=0.47\linewidth]{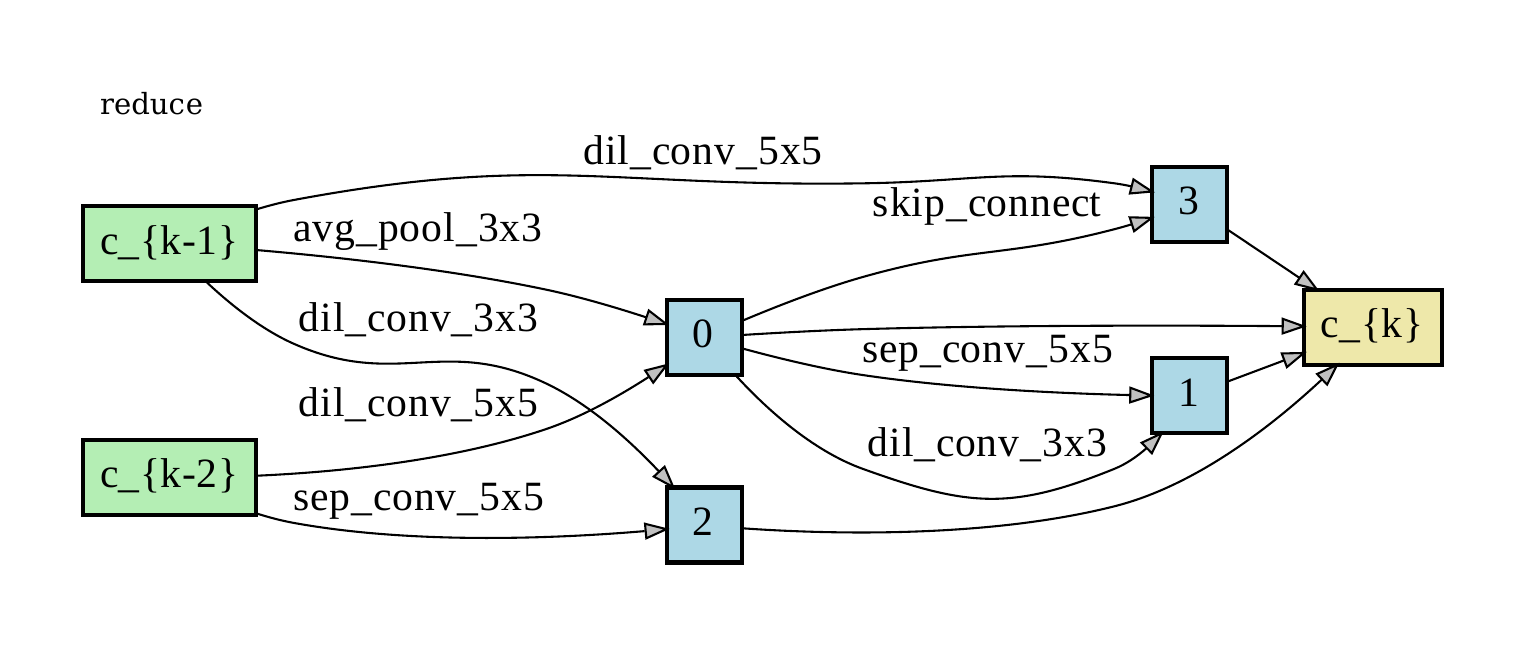}
}
\caption{The discovered cell architectures under the MiBB feature fault model. (a) Normal cell. (b) Reduction cell.}
\label{fig:discovered_f_cell}
\end{figure*}

As described in Sec.~\ref{sec:nas}, we conduct neural architecture searching without/with fault-tolerant training (i.e., FT-NAS and FTT-NAS, correspondingly). The per-MAC injection probability $p_m$ used in the search process is 1e-4. The reward coefficients $\alpha_r$ in Eq.~\ref{eq:reward} is set to $0.5$. In FTT-NAS, the loss coefficient $\alpha_l$ in Eq.~\ref{eq:loss} is also set to $0.5$. As the baselines for FT-NAS and FTT-NAS, we train ResNet-18, VGG-16, MobileNet-V2 with both normal training and FTT.  
For each model trained with FTT, we successively try per-MAC fault injection probability $p_m$ in \{3e-4, 1e-4, 3e-5\}, and use the largest injection probability with which the model could achieve a clean accuracy above 50\%. Consequently,
the ResNet-18 and VGG-16 are trained with a per-MAC fault injection probability of 1e-4 and 3e-5, respectively.

The discovered cell architectures are shown in Fig.~\ref{fig:discovered_f_cell}, and the evaluation results are shown in Table~\ref{table:NAS-f-results}. 
The discovered architecture \fftnet outperforms the baselines significantly at various fault ratios. In the meantime, compared with the most efficient baseline MobileNet-V2, the FLOPs number of \fftnet is comparable, and the parameter number is only 28.3\% (0.65M versus 2.30M).
If we require that the accuracy should be kept above $70\%$, MobileNet-V2 could function with a per-MAC error rate of 3e-6, and \fftnet could function with a per-MAC error rate larger than 1e-4. That is to say, while meeting the same accuracy requirements, \fftnet could function in an environment with a much higher SER.

We can see that FTT-NAS is much more effective than its degraded variant, FT-NAS. We conclude that, generally, NAS should be used in conjunction with FTT, as suggested by Eq.~\ref{eq:nas}. Another interesting fact is that, under the MiBB fault model, the relative rankings of the resilience capabilities of different architectures change after FTT: After FTT, MobileNet-V2 suffers from the smallest accuracy degradation among 3 baselines, whereas it is the most vulnerable one without FTT.

\subsection{Defend Against adSAF Weight Faults}
\label{sec:exp-weight}
We conduct FT-NAS and FTT-NAS under the adSAF model. The overall SAF ratio $p= p_0 + p_1$ is set to 8\%, in which the proportion of SAF0 and SAF1 is 83.7\% and 16.3\%, respectively (\revise{SAF0 ratio} $p_0$=6.7\%, \revise{SAF1 ratio} $p_1$=1.3\%). The reward coefficient $\alpha_r$ is set to $0.2$. The loss coefficient $\alpha_l$ in FTT-NAS is set to $0.7$. \revise{After the fault-tolerant training of the discovered architecture, we test the model accuracy with various SAF ratios (from 4\% to 12\%), while the relative ratio $p_0/p_1$ remains unchanged according to the numbers reported by \citet{chen2015rramdefect}.} 

\begin{figure*}[ht]
\centering
\subfloat[]{\includegraphics[width=0.46\linewidth]{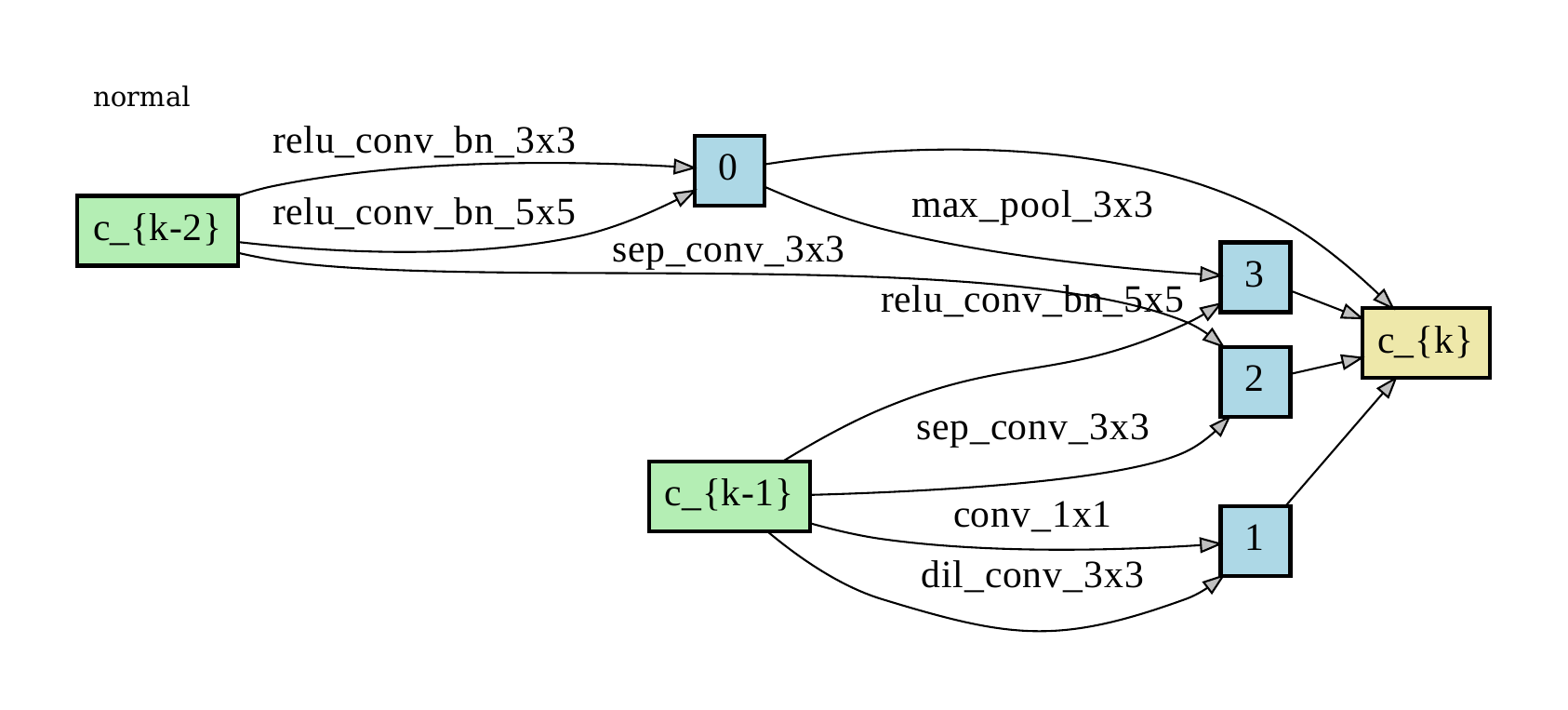}%
}
\hfill
\subfloat[]{\includegraphics[width=0.48\linewidth]{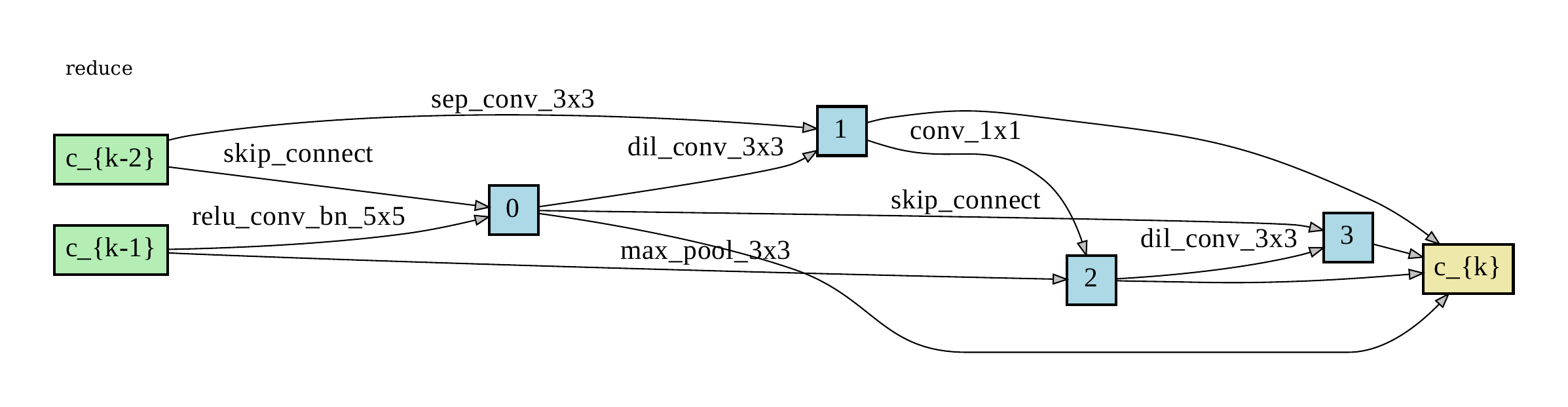} %
}
\caption{The discovered cell architectures under the adSAF weight fault model. (a) Normal cell. (b) Reduction cell.}
\label{fig:discovered_w_cell}
\end{figure*}

The discovered cell architectures are shown in Fig.~\ref{fig:discovered_w_cell}. As shown in Table~\ref{table:NAS-w-results}, 
the discovered \wftnet outperforms the baselines significantly at various test SAF ratios, with comparable FLOPs and less parameter number. We then apply channel augmentation to the discovered architecture to explore the performance of the model at different scales. We can see that models with larger capacity have better reliability under the adSAF weight fault model, e.g., $54.2\%$ (\wftnet-40) VS. $38.4\%$ (\wftnet-20) with 10\% adSAF faults.

\begin{table*}[ht]
\centering
\caption{Comparison of different architectures under the adSAF weight fault model}
\label{table:NAS-w-results}
\begin{tabular}{cc||ccccccccc}
\hline
\multirow{2}{*}{Arch}  & \multirow{2}{*}{Training} & \multirow{2}{*}{\begin{tabular}[c]{@{}c@{}}Clean\\ accuracy\end{tabular}} & \multicolumn{5}{c}{Accuracy (\%) with weight faults}    & \multirow{2}{*}{\#FLOPs} & \multirow{2}{*}{\#Params} \\ \cline{4-8}
                        & & &0.04 & 0.06&0.08 &0.10 &0.12     &                          &                           \\ \hline
ResNet-18           & clean     &{\bf 94.7} & {\bf 64.8}&	{\bf 34.9} &	17.8&	12.4&	11.0 & 1110M & 11.16M   \\
VGG-16              & clean     &93.1 &45.7&	21.7&	14.3&	12.6&10.6&626M&14.65M\\
MobileNet-V2        & clean     &92.3 &26.2&	14.3&	11.7&10.3&	10.5  & 182M  & 2.30M    \\\hline
\textbf{\wnet-20}      & clean     & 91.7 &   54.2 & 30.7	& {\bf 19.6}	&{\bf 15.5}	& {\bf 11.9} & 1020M & 3.05M    \\
\hline \hline
ResNet-18 & $p$=0.08     &    92.0   & 86.4	&77.9&	60.8&	41.6&	25.6   & 1110M & 11.16M    \\
VGG-16    & $p$=0.08     &    91.1   &  82.6&	73.3&	58.5&	41.7&	28.1 & 626M  & 14.65M    \\                        
MobileNet-V2 & $p$=0.08         &  86.3 & 76.6&	55.9&	35.7&	18.7&	15.1& 182M  & 2.30M    \\\hline
\textbf{\wftnet-20}$^\dagger$& $p$=0.08      &90.8   & 86.2&	79.5&	69.6&	53.5&	38.4 &  919M     &2.71M \\
\textbf{\wftnet-40}& $p$=0.08      &{\bf 92.1}   & {\bf 88.8}&	{\bf 85.5}&	 {\bf 79.3}&	{\bf 69.2} & {\bf 54.2}&	3655M     &10.78M \\\hline
\end{tabular}
\begin{minipage}{0.8\textwidth}
{\small
$\dagger$: The ``-$N$'' suffix means that the base of the channel number is $N$.
}
\end{minipage}
\end{table*}

\begin{figure*}[ht]
\begin{center}
\includegraphics[width=0.95\linewidth]{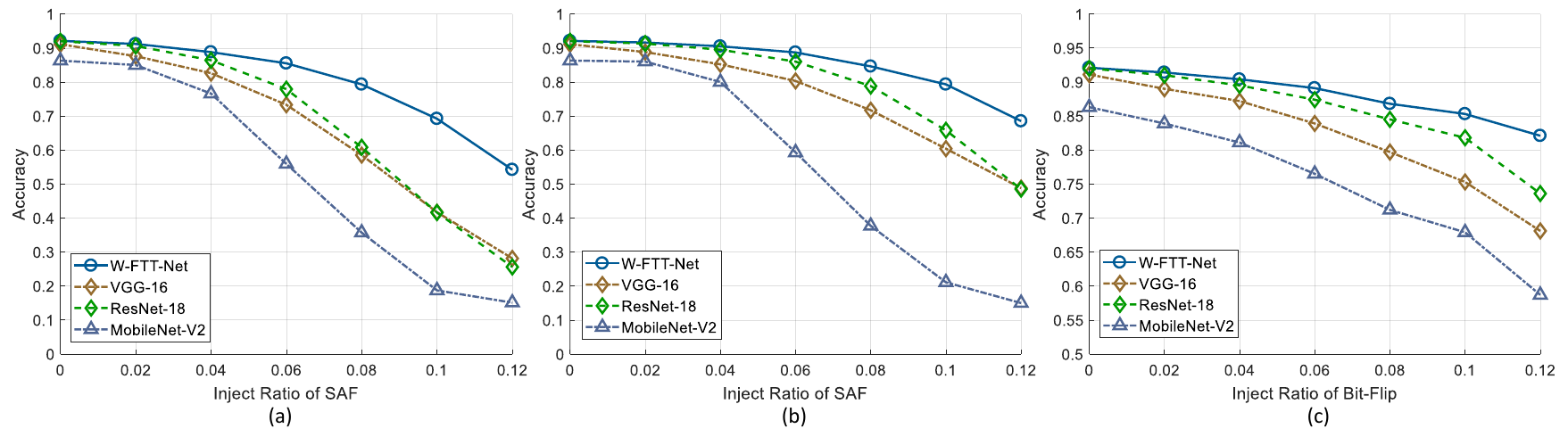}
\caption{Accuracy curves under different weight fault models. (a) W-FTT-Net under 8bit-adSAF model. (b) W-FTT-Net under 1bit-adSAF model. (c) W-FTT-Net under iBF model.}
\label{fig:weight-transfer}
\end{center}
\end{figure*}

In order to investigate whether the model FTT-trained under the adSAF fault model can tolerate other types of weight faults, we evaluate the reliability of \wftnet under 1bit-adSAF model and the iBF model. As shown in Fig.~\ref{fig:weight-transfer} (b)(c), under the 1bit-adSAF and iBF weight fault model, \wftnet outperforms all the baselines consistently at different noise levels. 

\subsection{The Effectiveness of The Learned Controller}
\label{sec:exp-controller}

To demonstrate the effectiveness of the learned controller, we compare the performance of the architectures sampled by the controller, with the performance of the architectures random sampled from the search space. For both the MiBB feature fault model and the adSAF weight fault model, we random sample $5$ architectures from the search space, and train them with FTT for 100 epochs. A per-MAC fault injection probability of 3e-4 is used for feature faults, and an SAF ratio of $8\%$ ($p_0$=6.7\%, $p_1$=1.3\%) is used for weight faults.

As shown in Table~\ref{table:random_sample_f} and Table~\ref{table:random_sample_w}, the performance of different architectures in the search space varies a lot, and the architectures sampled by the learned controllers, \fftnet and \wftnet, outperform all the random sampled architectures. Note that, as we use different preprocess operations for feature faults and weight faults (ReLU-Conv-BN 3x3 and SepConv 3x3, respectively), there exist differences in FLOPs and parameter number even with the same cell architectures.

\begin{table}[ht]
\centering
\caption{RNN controller VS. random samples under the MiBB feature fault model}
\label{table:random_sample_f}
\begin{tabular}{c|cccc}
\hline
     Model         & clean acc     & $p_m$=3e-4 & \#FLOPs & \#Params \\ \hline
sample1         & 60.2 &    19.5    & 281M & 0.81M    \\
sample2         & 79.7 &    29.7    & 206M & 0.58M    \\
sample3         & 25.0 &    32.2    & 340M & 1.09M    \\
sample4         & 32.9 &    25.8    & 387M & 1.23M    \\
sample5         & 17.4 &    10.8    & 253M & 0.77M    \\
\textbf{\fftnet}   & \textbf{88.6} &    \textbf{51.0}    & 245M & 0.65M    \\\hline
\end{tabular}
\end{table}

\begin{table}[ht]
\centering
\caption{RNN controller VS. random sample under the adSAF weight fault model}
\label{table:random_sample_w}
\begin{tabular}{c|cccc}
\hline
  Model            & clean acc     & $p$=8\% & \#FLOPs & \#Params \\ \hline
sample1         & 90.7 &    63.6    & 705M & 1.89M    \\
sample2         & 84.7 & 36.7       & 591M & 1.54M    \\
sample3         & 90.3 & 60.3       & 799M & 2.33M    \\
sample4         & 90.5 & 64.0       & 874M & 2.55M    \\
sample5         & 85.2 & 45.6       & 665M & 1.83M    \\
\textbf{\wftnet}   & \textbf{90.7} &    \textbf{68.5}    & 919M & 2.71M    \\ \hline
\end{tabular}
\end{table}

\subsection{Inspection of the Discovered Architectures}
\label{sec:inspection}

\noindent\textbf{Feature faults}: From the discovered cell architectures shown in Fig.~\ref{fig:discovered_f_cell}, we can observe that the controller obviously prefers SepConv and DilConv blocks over Relu-Conv-BN blocks. This observation is consistent with our anticipation. \textit{As under the MiBB feature fault model, operations with smaller FLOPs will result in a lower equivalent fault rate in the OFM.}

\textit{Under the MiBB feature fault model, there is a tradeoff between the capacity of the model and the feature error rate.} 
As the number of channels increases, the operations become more expressive, but the equivalent error rates in the OFMs also get higher. Thus there exists a tradeoff point of $c^*$ for the number of channels. Intuitively, $c^*$ depends on the per-MAC error rate $p_m$, the larger the $p_m$ is, the smaller the $c^*$ is.

\textit{Besides the choices of primitives, the connection pattern and combination of different primitives also play a role in making the architecture fault-tolerant.} To verify this, first, we conduct a simple experiment to confirm the preference of primitives: For each of the 4 different primitives (SepConv 3x3, SepConv 5x5, DilConv 3x3, DilConv 5x5), we stack 5 layers of the primitives, get the performance of the stacked NN after FTT training it with $p_m$=3e-4. The stacked NNs 
achieve the accuracy of 60.0\%, 65.1\%, 50.0\% and 56.3\% with $p_m=$1e-4, respectively. The stacked NN of SepConv 5x5 blocks achieves the best performance, which is of no surprise since the most frequent block in \fftnet is SepConv5x5. Then, we construct six architectures by random sampling five architectures with only SepConv5x5 connections and replacing all the primitives in \fftnet with SepConv 5x5 blocks. The best result achieved by these six architecture is 77.5\% with $p_m=$1e-4 (versus 86.2\% achieved by \fftnet). These illustrative experiments indicate that the connection pattern and combination of different primitives all play a role in the fault resilience capability of a neural network architecture. \\


\noindent\textbf{Weight faults}: Under the adSAF fault model, the controller prefers ReLU-Conv-BN blocks over SepConv and DilConv blocks. This preference is not so easy to anticipate.
We hypothesise that the weight distribution of different primitives might lead to different behaviors when encountering SAF faults. For example, if the quantization range of a weight value is larger, the value deviation caused by an SAF1 fault would be larger, and we know that a large increase in the magnitude of weights would damage the performance severely~\cite{hacene2019training}. We conduct a simple experiment to verify this hypothesis: We stack several blocks to construct a network, and in each block, one of the three operations (a SepConv3x3 block, a ReLU-Conv-BN 3x3 block, and a ReLU-Conv-BN 1x1 block) is randomly picked in every training step. The SepConv 3x3 block is constructed with a DepthwiseConv 3x3 and two Conv 1x1, and the ReLU-Conv-BN 3x3 and ReLU-Conv-BN 1x1 contain a Conv 3x3 and a Conv 1x1, respectively. 
After training, the weight magnitude ranges of Conv 3x3, Conv 1x1, and DepthwiseConv 3x3 are 0.036$\pm$0.043, 0.112$\pm$0.121, 0.140$\pm$0.094, respectively. \textit{Since the magnitude of the weights in 3x3 convolutions is smaller than that of the 1x1 convolutions and the depthwise convolutions, SAF weight faults would cause larger weight deviations in a SepConv or DilConv block than in a ReLU-Conv-BN 3x3 block.}

\section{Discussion}
\label{sec:discussion}

\subsection{Orthogonality}
Most of the previous methods are exploiting the inherent fault resilience capability of existing NN architectures to tolerate different types of hardware faults. In contrast, our methods improve the inherent fault resilience capability of NN models, thus effectively increase the algorithmic fault resilience ``budget'' to be utilized by hardware-specific methods. Our methods are orthogonal to existing fault-tolerance methods, and can be easily integrated with them, e.g., 
helping hardware-based methods to reduce the overhead largely.

\subsection{Data representation}
In our work, an 8-bit dynamic fixed-point representation is used for the weights and features. As pointed out in Sec.~\ref{sec:inspection}, the dynamic range has impacts on the resilience characteristics against weight faults. The data format itself obviously decides or affects the data range. \citet{yan2019whense} found out that the errors in exponent bits of the 32bit floating-point weights have large impacts on the performance. \citet{li2017understanding} investigated the resilience characteristics of several floating-point and non-dynamic fixed-point representations.

\subsection{Other search strategies (controllers)}
\revise{A NAS system mainly consists of three components that work together: The search space that defines the architectural decisions to make; The evaluator that assesses the performance of an architecture; The search strategy (controller) that explores the search space using the rewards provided by the evaluator. Apart from few exceptions such as the differentiable NAS methods~\cite{DARTS}, the NAS system could be built in a modularized and decoupled way, in which the controller and the evaluator can be designed independently. Our major modifications to the NAS framework lie in that 1) We analyze different categories of faults existing in various devices, and conduct fault insertion with the formulated fault model when \textbf{evaluating the reward}. 2) The fault-tolerant training techniques should be incorporated when \textbf{training the supernet weights}, to reduce the gap between the search and final training stages, since it is a common technique for training a fault-tolerant NN model. Note that these two amendments are all on the evaluator component. Despite that we choose to use the popular reinforcement learning-based controller, other controllers such as evolutionary-based~\cite{real2019aging} and predictor-based ones~\cite{ning2020generic} could be incorporated into the FTT-NAS framework easily. The application of other controllers is outside the scope of this work's interests.}

\subsection{Fault model}\label{sec:discuss_fault_model}
\noindent\textbf{Limitation of application-level fault model}: There are faults that are hard or unlikely to model and mitigate by our methods, e.g., timing errors, routing/DSP errors in FPGA, etc. A hardware-in-the-loop framework could be established for a thorough evaluation of the system-level fault hazards. Anyway, since the correspondence between these faults and the application-level elements are subtle, it's more suitable to mitigate these faults in the lower abstraction layer.\\

\noindent\textbf{\revise{FPGA platform}}: 
In the MiBB feature fault model, we assume that the add operations are spatially expanded onto independent hardware adders, which applies to the template-based designs~\cite{venieris2017convnet}. For ISA (Instruction Set Architecture) based accelerators~\cite{qiu2016going}, the NN computations are orchestrated using instructions, time-multiplexed onto hardware units. In this case, the accumulation of the faults follows a different model and might show different preferences among architectures. Anyway, the FTT-NAS framework is general and could be used with different fault models. We leave the exploration and experiments of other fault models for future work.\\

\noindent\textbf{\revise{RRAM platform}}: 
\revise{As for the RRAM platform, this paper mainly focuses on discovering fault-tolerant neural architecture to mitigate SAFs, which have significant impacts on computing accuracy. In addition to SAFs, the variation is another typical RRAM non-ideal factor that may lead to inaccurate computation. There exist various circuit-level optimizations that can mitigate the computation error caused by the RRAM variation.
Firstly, with the development of the RRAM device technology, a large on/off ratio of RRAM devices (i.e., the resistance ratio of the high resistance state and the low resistance state) can be obtained (e.g., $10^3$~\cite{Woo2020ExploitingDR}). And a large on/off ratio makes the bit line current difference among different computation results obvious and thus improves the fault tolerance capability against variation. Secondly, in existing RRAM-based accelerators, the number of activated RRAM rows at one time is limited. For example, only four rows are activated in each cycle, which provides a sufficient signal margin against process variation~\cite{xue2021a22nm}. In contrast, compared with the process variation, it is more costly to mitigate SAFs by circuit-level optimization (e.g., existing work utilizes costly redundant hardware to tolerate SAFs~\cite{huangfu2017computation}). Thus, we aim at tolerating the SAFs from the algorithmic perspective. Anyway, simulating the variation is a meaningful extension of the general FTT-NAS framework for the RRAM platform, and we leave it for future work.}\\

\noindent\textbf{Combining multiple fault models}: \revise{We experiment with a fault model at a time, and do not combine different fault models. And our experimental results show that the architectural preferences of the adSAF and iBB feature fault models are distinct (See discussions in Sec.~\ref{sec:inspection}). Fortunately, the two types of faults that we experiment with would not co-exist for the same part of an NN model: IBB feature faults (caused by FPGA LUTs errors) and the adSAF weight faults (in the RRAM crossbar).
Nevertheless, there indeed exist scenarios that some weights and feature errors could happen simultaneously on one platform. For example, iBF in the feature buffer and SAF in the crossbar can occur simultaneously in an RRAM-based accelerator.
However, on the same platform in the same environment, the influences of different types of errors would usually be vastly different. For example, compared with the accuracy degradation caused by the SAF errors in the RRAM cell, the influences of iBF errors in the feature buffer could usually be ignored on the same device. }

\revise{As a future direction, it might be interesting to combine these fault models to search for a neural architecture to be partitioned and deployed onto a heterogeneous hardware system. In that case, the fault patterns, along with the computation and memory access patterns of multiple platforms should be considered jointly.}\\

\section{Conclusion}
\label{sec:conclusion}
In this paper, we analyze the possible faults in various types of NN accelerators and formalize the statistical fault models from the algorithmic perspective. After the analysis, the MAC-i.i.d Bit-Bias (MiBB) model and the arbitrary-distributed Stuck-at-Fault (adSAF) model are adopted in the neural architecture search for tolerating feature faults and weight faults, respectively. To search for the fault-tolerant neural network architectures, we propose the multi-objective Fault-Tolerant NAS (FT-NAS) and Fault-Tolerant Training NAS (FTT-NAS) method. In FTT-NAS, the NAS technique is employed in conjunction with the Fault-Tolerant Training (FTT). The fault resilience capabilities of the discovered architectures, \fftnet and \wftnet, outperform multiple manually designed architecture baselines, with comparable or fewer FLOPs and parameters. And \wftnet trained under the 8bit-adSAF model can defend against other types of weight faults. Generally, compared with FT-NAS, FTT-NAS is more effective and should be used. In addition, through the inspection of the discovered architectures, we find that since operation primitives differ in their MACs, expressiveness, weight distributions, they exhibit different resilience capabilities under different fault models. The connection pattern is also shown to have influences on the fault resilience capability of NN models.

\begin{acks}
This work was supported by National Natural Science Foundation of China (No. U19B2019, 61832007, 61621091), National Key R\&D Program of China (No. 2017YFA02077600); Beijing National Research Center for Information Science and Technology (BNRist); Beijing Innovation Center for Future Chips; the project of Tsinghua University and Toyota Joint Research Center for AI Technology of Automated Vehicle (TT2020-01); Beijing Academy of Artificial Intelligence.
\end{acks}

\bibliographystyle{ACM-Reference-Format}
\bibliography{sample-base}










\end{document}